\documentclass[article,twocolumn,superscriptaddress,floatfix]{revtex4-2}
\usepackage{tabularx}
\usepackage{graphicx}
\usepackage{amsmath}
\usepackage{xcolor}

\graphicspath{ {./Figures/} }

\newcolumntype{Y}{>{\centering\arraybackslash}X}
\newcolumntype{P}[1]{>{\centering\arraybackslash}p{#1}}

\begin{document}
    \title{Quantum Coherent Transport of 1D Ballistic States in Second-Order Topological Insulator Bi$_4$Br$_4$}
    
    \author{Jules Lefeuvre}
    \email{jules.lefeuvre@universite-paris-saclay.fr}
    \affiliation{Université Paris-Saclay, CNRS, Laboratoire de Physique des Solides, 91405, Orsay, France.}
    \author{Masaru Kobayashi}
    \affiliation{Materials and Structures Laboratory, Institute of Science Tokyo, Yokohama, Japan.}
    \author{Gilles Patriarche}
    \author{Nathaniel Findling}
    \affiliation{Université Paris-Saclay, CNRS, Centre de Nanosciences et de Nanotechnologies, 91120, Palaiseau, France.}
    \author{David Troadec}
    \affiliation{Université de Lille, CNRS, Institut d'Electronique, de Microélectronique et de Nanotechnologie,  59652, Villeneuve d'Ascq, France}
    \author{Meydi Ferrier}
    \author{Sophie Guéron}
    \author{Hélène Bouchiat}
    \affiliation{Université Paris-Saclay, CNRS, Laboratoire de Physique des Solides, 91405, Orsay, France.}
    \author{Takao Sasagawa}
    \affiliation{Materials and Structures Laboratory, Institute of Science Tokyo, Yokohama, Japan.}
    \author{Richard Deblock}
    \email{richard.deblock@universite-paris-saclay.fr}
    \affiliation{Université Paris-Saclay, CNRS, Laboratoire de Physique des Solides, 91405, Orsay, France.}
    
    % \date{\today}
    
    \begin{abstract}
        We investigate quantum transport in micrometer-sized single crystals of Bi$_4$Br$_4$, a material predicted to be a second-order topological insulator. 1D topological states with long phase coherence times are revealed via the modulation of quantum {\color{black}interference} with magnetic field and gate voltage. In particular, we demonstrate the existence of Aharonov-Bohm interference between 1D ballistic states several micrometers long, that we identify as phase-coherent hinge modes on neighboring step edges at the crystal surface. These Aharonov-Bohm {\color{black}oscillations} are made possible by a disordered phase-coherent contact region, the existence of which is confirmed by scanning transmission electron microscopy combined with energy-dispersive X-ray spectroscopy (STEM-EDX) of FIB lamellae. Their coherent nature modulates the transmission of the 1D edge states, leading to weak antilocalization and universal conductance fluctuations with surprisingly large characteristic fields and a strongly anisotropic behavior. 
    	These complementary experimental results provide a comprehensive, coherent description of quantum transport in Bi$_4$Br$_4$, and establish the material as {\color{black}a second-order topological insulator} with topologically protected 1D ballistic states.
    \end{abstract}

	\maketitle
		
%%%%%%%%%%%%%%%%%%%%%% Intro %%%%%%%%%%%%%%%%%%%%%%%%%%%%%%%%%%	
	\section{Introduction}
	    An important recent achievement in condensed matter physics is the discovery of a new class of materials, topological insulators (TIs), whose bulk is insulating, while the boundaries are conducting \cite{Hasan2010,Qi2011,Ando2013}. In particular, the 1D edges of 2DTIs realize the quantum spin hall (QSH) state \cite{Kane2005,Bernevig2006}, where current is carried dissipationlessly by two counterpropagating ballistic channels with opposite spins, forming a helical edge state. This opens many possibilities, ranging from dissipationless charge and spin transport at room temperature, to new avenues for quantum computing. Among the recent additions to the TI classification, second order topological insulators (SOTIs) \cite{Shindler2018,Xie2021} are $n$-dimensional crystals exhibiting $(n-2)$-dimensional topological states. In particular, 3D SOTIs have insulating bulk and surfaces, but topologically protected, one-dimensional helical “hinge” states. 1D helical states have been evidenced in several systems in 2D such as HgTe quantum wells \cite{Konig2007}, monolayer WTe$_2$ \cite{Fei2017,Wu2018}, or TaIrTe$_4$ \cite{Tang2024} and in 3D systems such as crystalline bismuth \cite{Murani2017,Schindler2018b}. Unfortunately, the expected topological protection has turned out to be less robust than anticipated, notably due to the existence of conduction in the bulk and surface, or charge puddles \cite{Weber2024}. This complicates the fundamental study of the edge states, and motivates the search for different TIs with a reduced contribution of the nontopological bulk states. Among newly discovered TIs, Bi$_4$Br$_4$ appears to be a very promising material, with a large bulk gap ($\gtrsim 200$ meV) \cite{Zhou2014,Zhou2015,Hsu2019,Tang2019,Yoon2020,Han2022,Lin2024}, and strong experimental indications of a SOTI character \cite{Noguchi2021,Yang2022,Zhong2025,Zhao2023} .
	    
\begin{figure*}[tb]
    		\centering
    		\includegraphics[width=\textwidth]{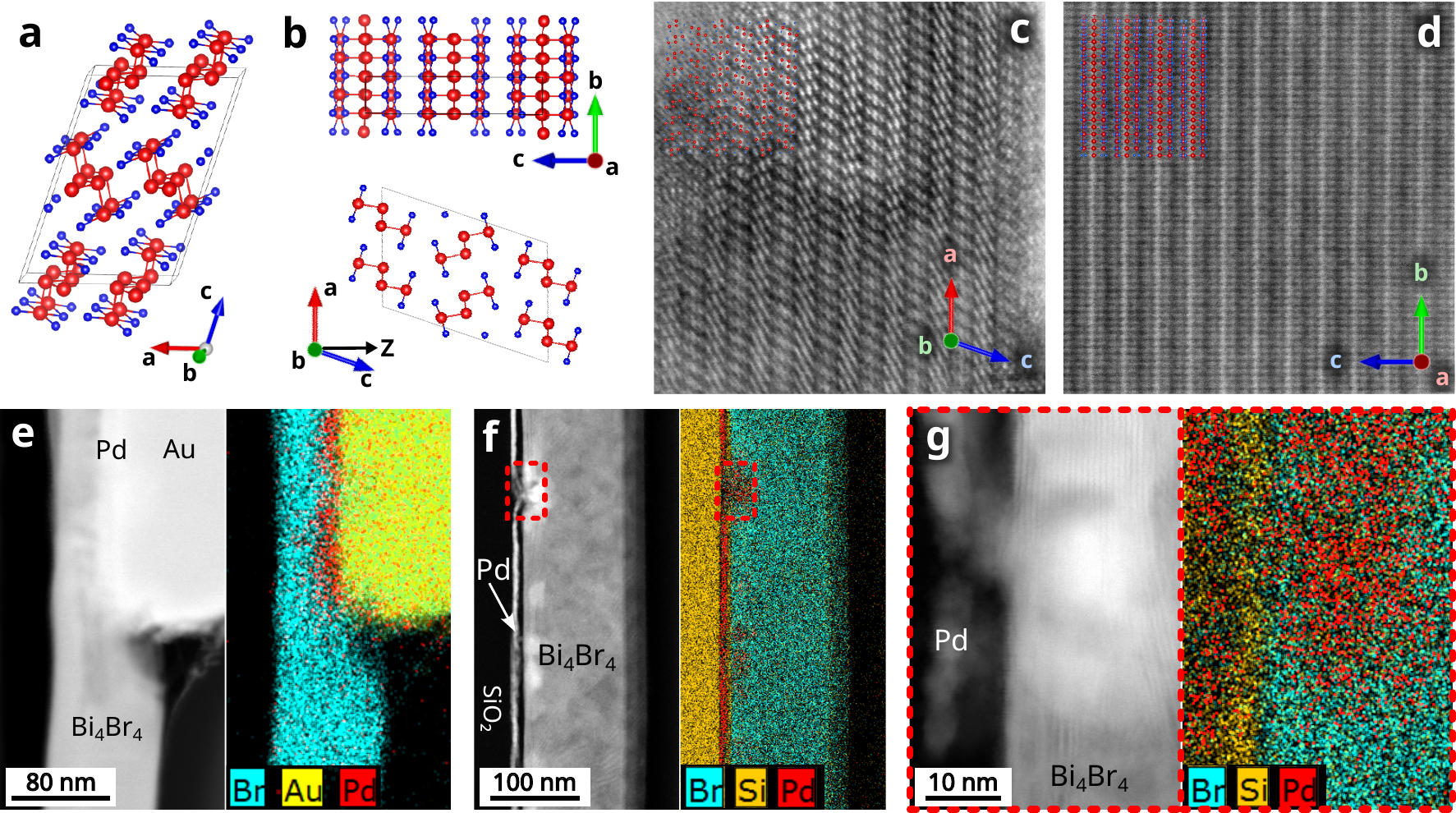}
    		\caption{Structure of Bi$_4$Br$_4$ crystals far from and in the contact region. (a) Crystalline structure of Bi$_4$Br$_4$ and crystal orientation. Bi atoms are figured in red, {\color{black}and} Br in blue. (b) Structure of Bi$_4$Br$_4$ along the $a$ and $b$ axis. The Z direction, perpendicular to the (a,b) plane of the flakes, is indicated. (c) High-resolution high-angle-annular dark-field (HAADF) STEM images of FIB {\color{black}lamellae} of contacted Bi$_4$Br$_4$ flakes obtained along the $b$ direction{\color{black},} away from the contact area. The expected crystalline {\color{black}structure is} superimposed on the picture. The agreement between theoretical and real atoms positions is excellent. (d) Similar image along the $a$ direction. (e) Contact region of Au/Pd (100 nm/7 nm) top-contacted electrodes, showing crystal damage both under and away from the electrode. The EDX analysis (right image) shows a diffusion of Bi in the Pd and, to a lesser extent, in the Au layers. The spurious Si signal in the gold electrode is due to the proximity of a fluorescence band of Au with the Si K$\alpha$. (f) Contact region for 7-nm-thick Pd electrodes placed below the Bi$_4$Br$_4$ flakes. Diffusion of Pd in the crystal is also visible. (g) Enlargement of the area denoted by the red dashed rectangle in (f), which evidence disordered contact regions of size around 100 nm.}
    		\label{fig:EDX}
        \end{figure*}

	    This material, which in its $\alpha$ form has a bilayer structure of chains oriented along the crystal's $b$ direction [Fig. \ref{fig:EDX}(a) and \ref{fig:EDX}(b)], has indeed been predicted to show the QSH state for the monolayer \cite{Zhou2014,Zhou2015} and {\color{black}the multilayer has been classified as a SOTI} \cite{Hsu2019,Tang2019,Yoon2020,Han2022,Lin2024,Han2023,Shumiya2022,Hofman2024,Yang2022,Peng2021}. The SOTI {\color{black}character} in the multilayer case results from the coupling of helical edge states upon stacking of 2D monolayers, leading to gapped side surfaces with a gap of the order of 30 meV. On a perfect crystal, only two helical states {\color{black}located} at hinges remain uncoupled, {\color{black}which} location depends on the parity of the number of layers. 
	    
	    The band structure of Bi$_4$Br$_4$ has been probed by angle-resolved photoemission spectroscopy (ARPES) \cite{Noguchi2021,Yang2022,Zhong2025} and micro-ARPES \cite{Zhao2023}. This latter work confirmed the existence of gapped lateral surfaces, with a gap ranging from 28 to 40 meV, and resolved their spin texture, indicating a spin polarization perpendicular to the direction of the chains. {\color{black}In addition,} optical pump–probe microscopy at room temperature revealed a relatively long carrier lifetime for boundary states attributed to their linear dispersion relation \cite{Han2023}. {\color{black}Lastly,} conducting hinge states have been probed by STM \cite{Shumiya2022,Hofman2024,Yang2022,Peng2021} up to room temperature, on mono- and few-layers-high step edges. These states are gapped by a magnetic field perpendicular to chains with a Landé factor of 53, indicating a spin quantization axis along the $b$ axis, in stark contrast to the spin polarization of bulk \cite{Lin2024} and lateral surface states \cite{Zhao2023}.
	    
	    With regard to electronic transport, experiments carried out on macroscopic samples indicate that the temperature dependence of resistivity is close to that of a semiconductor below 50 K and often shows a local maximum around 200 K \cite{Benda1978,Filatova2007,Li2019,Noguchi2021,Chen2022,Yang2022,Gong2024}. It has been proposed that the former results from either competition between bulk and hinge states \cite{Noguchi2021,Chen2022}, or the appearance of a charge density wave \cite{Filatova2007,Li2019} or a variable-range hopping induced by Br vacancies  \cite{Gong2024}{\color{black}, while} the 200 K feature is attributed to a Lifshitz transition \cite{Chen2022,Yang2022}. At low temperature, macroscopic samples exhibit weak localization or weak antilocalization depending on {\color{black}the field orientation} \cite{Zhong2022,Zhong2024}. High-pressure experiments found a metal-insulator transition that can even lead to a superconducting state at pressures above 3.8 GPa \cite{Li2019}. By probing micrometer-size crystals obtained by exfoliation, several temperature behaviors were obtained : a nearly temperature-independent resistivity for relatively long flakes \cite{Qiao2021}, a gate tunable resistance attributed to competition of bulk and surface conduction \cite{Wu2023}, or a thermal activation of bulk carriers \cite{Hossain2024}. This latter reference demonstrated an Aharonov-Bohm effect associated to quantum interference of coherent hinge states around single lateral surfaces in four- and six-layer flakes.
	    
	    In the present work, we study quantum transport in {\color{black}micrometer-sized} Bi$_4$Br$_4$ flakes with thicknesses ranging from 50 to 150 nm. We find evidence of phase-coherent ballistic 1D states over several microns via {\color{black}the} Aharonov-Bohm effect, as well as negative four-wire resistances. Interestingly, the samples also consistently display weak antilocalization and universal conductance fluctuations, which are staples of 2D coherent transport in the diffusive regime. These apparently conflicting observations can be reconciled by considering a {\color{black}small, coherent, but} disordered region between the pristine crystal and the contact, which enables both interhinge coherence and anomalous 2D quantum interference. The existence of this region is confirmed by direct imaging and elemental analysis using STEM-EDX. While this contact regime prevents observation of the quantification of conductance generally expected in QSH systems, it {\color{black}provides} extremely rich signatures of the underlying topological transport.

%%%%%%%%%%%%%%%%%%% Exfoliation and contacts %%%%%%%%%%%%%%%%%%%%%%%%%%%%%%%%%%%%%%			
	\section{Exfoliation and contacting of B\lowercase{i}$_4$B\lowercase{r}$_4$ flakes}
	    
	    The Bi$_4$Br$_4$ crystals were synthesized by the chemical vapor transport method \cite{Noguchi2019,Noguchi2021}. The samples were exfoliated in a glove box under argon atmosphere with O$_2<$0.2 ppm and H$_2$O$<$5 ppm. The exfoliation and transfer of micrometer-size Bi$_4$Br$_4$ flakes was done using PDMS stamps. We check with STEM imaging that the crystalline quality of the exfoliated flakes is preserved before any contacts is made (Sec. 1 in Supplemental Material \cite{SM}). 
    
	    In order to contact the flakes we developed two strategies. The first one is based on contacts made on top of the flake. This is done by e-beam lithography, with low-temperature processes to preserve the quality of the flakes (PMMA resist baking below 100\textcelsius), and metal deposition in an e-beam evaporator after argon ion beam etching.  Different contact materials were tried. The best results were obtained with palladium (Pd). The samples are thus contacted by Pd(7 nm)/Au(100 nm) electrodes; this leads to k$\Omega$ two-wire resistances at room temperature. However, this method damages the crystal under the contacts and exposes the flakes to organic solvents and ambient air during the e-beam lithography and lift-off processes. 
	    To avoid these problems, we also tried contacting the flakes by transferring them onto predefined metallic electrodes, fabricated by e-beam and optical lithography, with e-beam metal deposition. The Bi$_4$Br$_4$ flakes are then pressed onto the electrodes in Ar atmosphere. During this process, we found that heating the transfer stage to 80°C produced Ohmic contacts more reliably. The best results were obtained, as in the case of contacts on top, with palladium electrodes and also yields two-wire resistances in the k$\Omega$ range.
		
	    To understand the nature of the contact region between Pd electrodes and Bi$_4$Br$_4$ flakes, we produced FIB {\color{black}lamellae} of the region of the contacts, as well as slices away from the {\color{black}contacts} and analyzed them in a STEM. The results are presented in Fig. \ref{fig:EDX}. Sections taken far from the contacts show excellent crystalline quality, even after exposure to ambient air [Fig. \ref{fig:EDX}(c) and \ref{fig:EDX}(d)]. By contrast, for both contacting schemes, the contact regions are disordered, with Pd/Bi interdiffusion, and have a typical {\color{black}size of approximately 100 nm}.

%%%%%%%%%%%%%%%%%%% Anisotropy %%%%%%%%%%%%%%%%%%%%%%%%%%%%%%%%%%%%%%%%%%%%%	
	\section{Anisotropic electronic transport in B\lowercase{i}$_4$B\lowercase{r}$_4$ flakes}
		
		We have performed electronic transport {\color{black}measurements} on more than ten samples. In the following{\color{black},} we focus on four samples. One, denoted $\mathcal{L}$, is a flake with six contacts [top contacts with Pd(7 nm)/Au(100 nm) electrodes], thus defining five segments, with length ranging from 1 to 5 $\mu$m. The flake has a total length of 30 $\mu$m, a width of 2.1 $\mu$m and a thickness of around 100 nm [Fig. \ref{fig:interf}(a)]. The flake is deposited on a doped silicon substrate, which serves as a back gate, covered by 500 nm of oxide.
		\begin{figure}[tb]
    		\centering	
    		\includegraphics[width=\columnwidth]{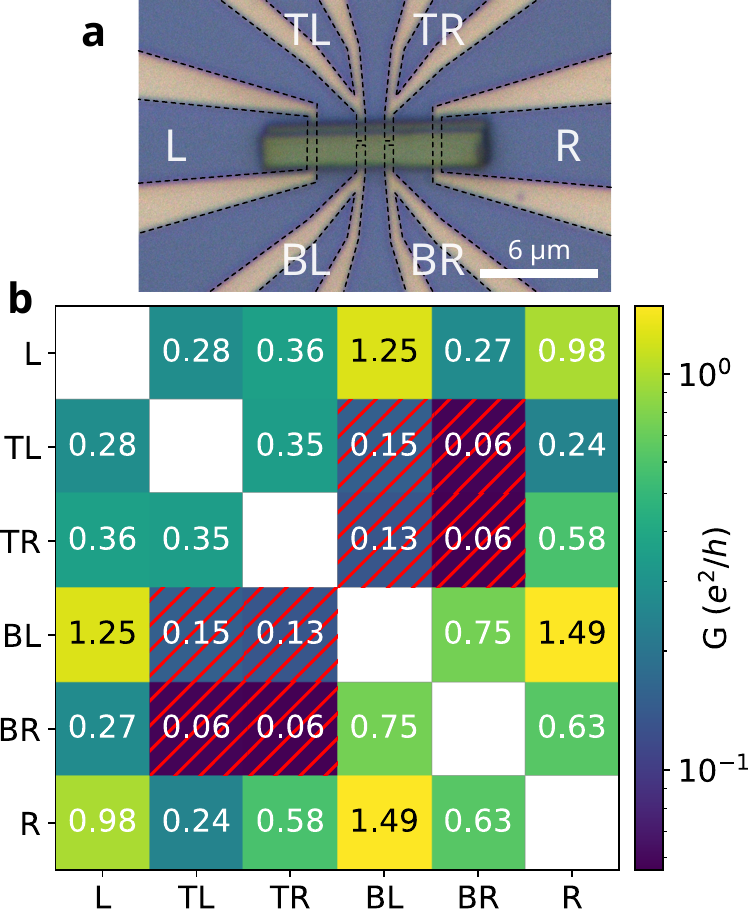}
    		\caption{Anisotropic transport in Bi$_4$Br$_4$ sample $H_1$ at 10 mK and {\color{black}under a 500 nA dc bias}. (a) Optical picture of sample $H_1$ with labeling of the electrodes. (b) Conductance matrix of sample $H_1$. The conductances are small in the transverse direction ($a$ axis of the crystal), highlighted by red hatching, and much higher, by a factor 10, in the longitudinal direction ($b$ axis of the crystal).}
    		\label{fig:R-net}
	    \end{figure}
	
		We also discuss three samples with a Hall bar geometry. Sample $\mathcal{H}_1$, shown in Fig. \ref{fig:R-net}(a), is a flake deposited on top of Ti(5 nm)/Au(10 nm)/Pd(5 nm) contacts. Sample $\mathcal{H}_3$ has similar contacts, while sample $\mathcal{H}_2$ has {\color{black}Pd(7 nm)/Au(100 nm)} contacts deposited from the top (Fig. \ref{fig:anisotropy} and Sec. 2 of Supplemental Material \cite{SM}).
		
\begin{figure}[tb]
    		\centering	\includegraphics[width=\columnwidth]{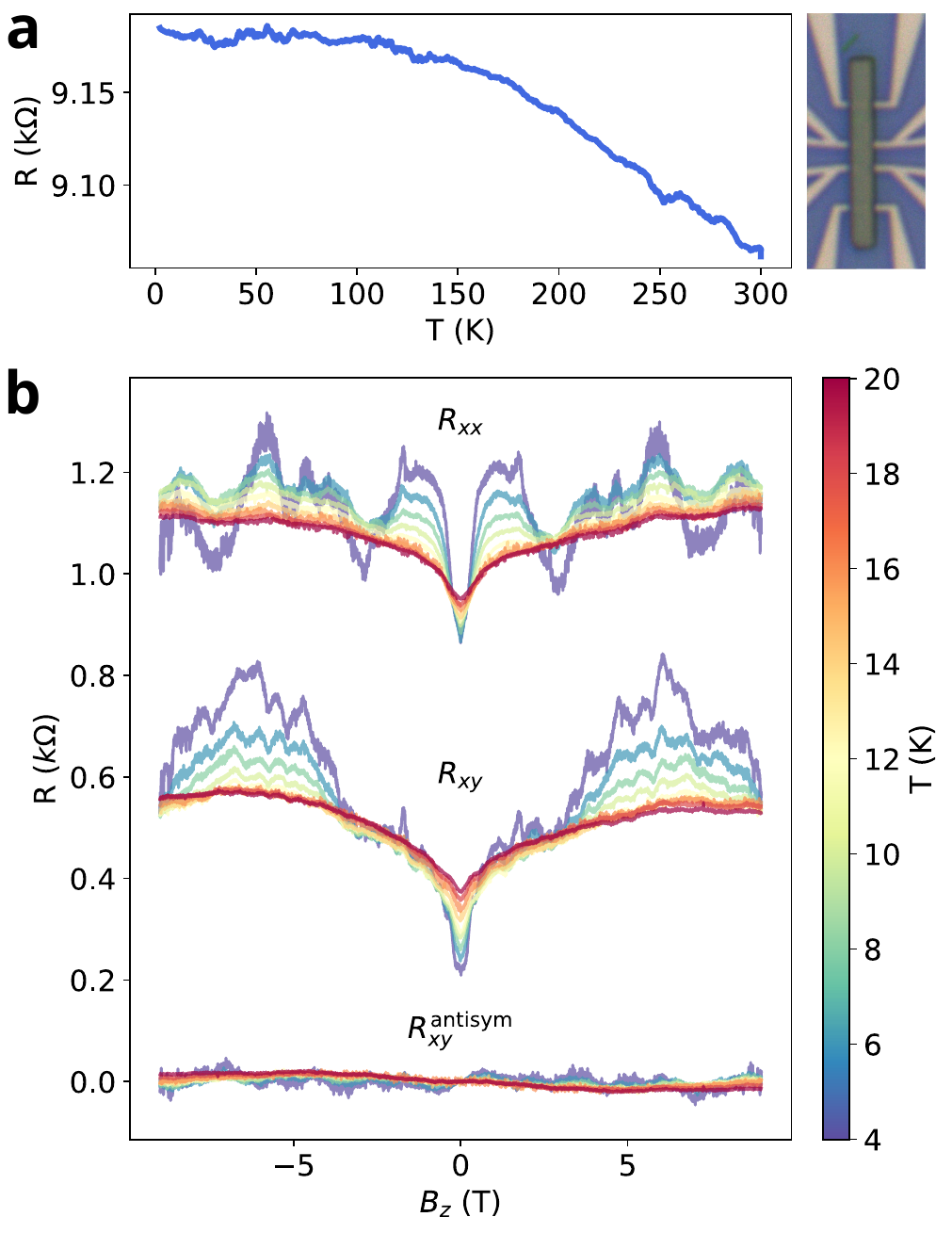}
    		\caption{Temperature dependence of electronic transport in Bi$_4$Br$_4$. (a) Temperature dependence of a representative Bi$_4$Br$_4$ sample, measured in a physical property measurement system. A Savitzky-Golay filter was applied to the data to get rid of the high-frequency noise. Right : optical picture of the sample used for temperature dependence. (b) Longitudinal ($R_{xx}$) and transverse ($R_{xy}$) magnetoresistance at temperature between 4 (purple) and 20 K (red) for sample $\mathcal{H}_1$. $R_{xy}^{\mathrm{antisym}}$ corresponds to the antisymmetric part of the transverse resistance and shows no Hall effect.}
    		\label{fig:R-hall}
	    \end{figure}

	    All samples display four-wire resistance values in the k$\Omega$ range at room temperature that do not change significantly down to 10 K [Fig. \ref{fig:R-hall}(a)]. There is a sharp increase of resistance below 1 K, which we attribute to the mesoscopic nature of our sample rather than to the band structure (Sec. 7 of Supplemental Material \cite{SM}). The absence of a strong temperature dependence in our measurements contrasts with other works \cite{Noguchi2021,Chen2022,Wu2023,Filatova2007,Li2019,Gong2024,Chen2022,Yang2022} and suggests that our samples are relatively clean, with a low density of doping defects, and, therefore, have a rather temperature-stable chemical potential. Moreover, the absence of a temperature-activated behavior across the whole temperature range suggests that the Fermi level lies at least 26 meV away from a bulk band edge, thus clearly in the bulk gap.

		The four-wire resistances of the segments of sample $\mathcal{L}$ do not increase proportionally to their lengths, which indicates that they are dominated by a contact resistance (Sec. 3 of Supplemental Material \cite{SM}). This can be explained by {\color{black}two} competing scenarios : either a resistive interface between the Pd/Au contacts and {\color{black}well-conducting} bulk or surface states, or ballistic conduction{\color{black}, likely due to hinge states}. The variations in resistance from segment to segment can, thus, be explained either by fluctuations in interface quality or by differences in the number of edge modes contacted due to insulating resist residues under the electrodes.
		
		In order to elucidate the flow of current inside the Hall bar devices, we use a formula derived in the context of graph theory \cite{Bapat2004,Devriendt2022}:
		    \begin{equation}
    			\mathcal{G} = -2\mathcal{R}^{-1} + 2 \frac{\mathcal{R}^{-1} \mathbf{u} \mathbf{u}^T \mathcal{R}^{-1}}{\mathbf{u} \mathcal{R}^{-1} \mathbf{u}^T}
    			\label{eq:resDis}
            \end{equation}
        which relates two quantities, the Laplacian matrix $\mathcal{G}$ and the resistance distance matrix $\mathcal{R}$, with $\mathbf{u}$ a vector with all coefficients equal to 1. Modeling the sample as a network of lumped resistances with nodes representing terminals, the resistance distance matrix $\mathcal{R}$ contains the two-wire resistances between each possible pair of terminals of the resistance network. The Laplacian matrix $\mathcal{G}$ has matrix elements $\mathcal{G}_{ij}=-G_{ij}$, with $G_{ij}$ the discrete node-to-node conductances, and diagonal coefficients ensuring current conservation $\mathcal{G}_{ii}=\sum_{j \neq i} G_{ij}$. $\mathcal{G}$, thus, corresponds to the conductance matrix in the context of mesoscopic transport \cite{Datta2009}. This result, therefore, provides a direct determination of the conductance matrix from simple resistance measurements \cite{reciprocalCond}.
        
		The conductance matrix of $\mathcal{H}_1$ is given in Fig. \ref{fig:R-net}(b) (see Sec. 2 in Supplemental Material \cite{SM} for samples $\mathcal{H}_2$ and $\mathcal{H}_3$). All contacts to the flake are doubled, so that the lead resistance can be removed for every two-wire configuration. We find a very anisotropic transport, with relatively high conductances, of the order of the conductance quantum, along the $b$ direction of the crystal. Noticeably, the conduction along the $a$ direction is consistently among the smallest in all of our devices, with conductances {\color{black}under} 0.15 $e^2/h$ for contacts 500 nm apart for $\mathcal{H}_1$. Nonzero transverse conductances may result from residual interface resistances that are not eliminated by our measurement procedure.
	
        Finally, we have carried out magnetotransport measurements (Fig. \ref{fig:R-hall}) on the same sample, which revealed an essentially symmetrical four-wire transverse resistance dominated by quantum interferences at low temperature (see the next section). The antisymmetrical fraction of the contribution to the magnetoresistance is extremely small, with no clear linear dependence in magnetic field, indicating the absence of Hall effect in the sample. {\color{black} Accordingly, the longitudinal resistance exhibits no Shubnikov-de Haas oscillations (Sec. 8 in Supplemental Material \cite{SM}), further demonstrating the absence of transverse transport.}
        
        These results are a striking demonstration that the bulk and (001) surfaces are insulating. It suggests that conduction is {\color{black}either} due to {\color{black}the lateral (100)} surfaces or {\color{black}to 1D states} along the $b$ direction. 

%%%%%%%%%%%%%%%%%%% Quantum Transport %%%%%%%%%%%%%%%%%%%%%%%%%%%%%%%%%%%%%%%%%%%%%	
	\section{Low-temperature quantum transport}
	
\begin{figure*}[tb]
		\centering
		\includegraphics[width=\textwidth]{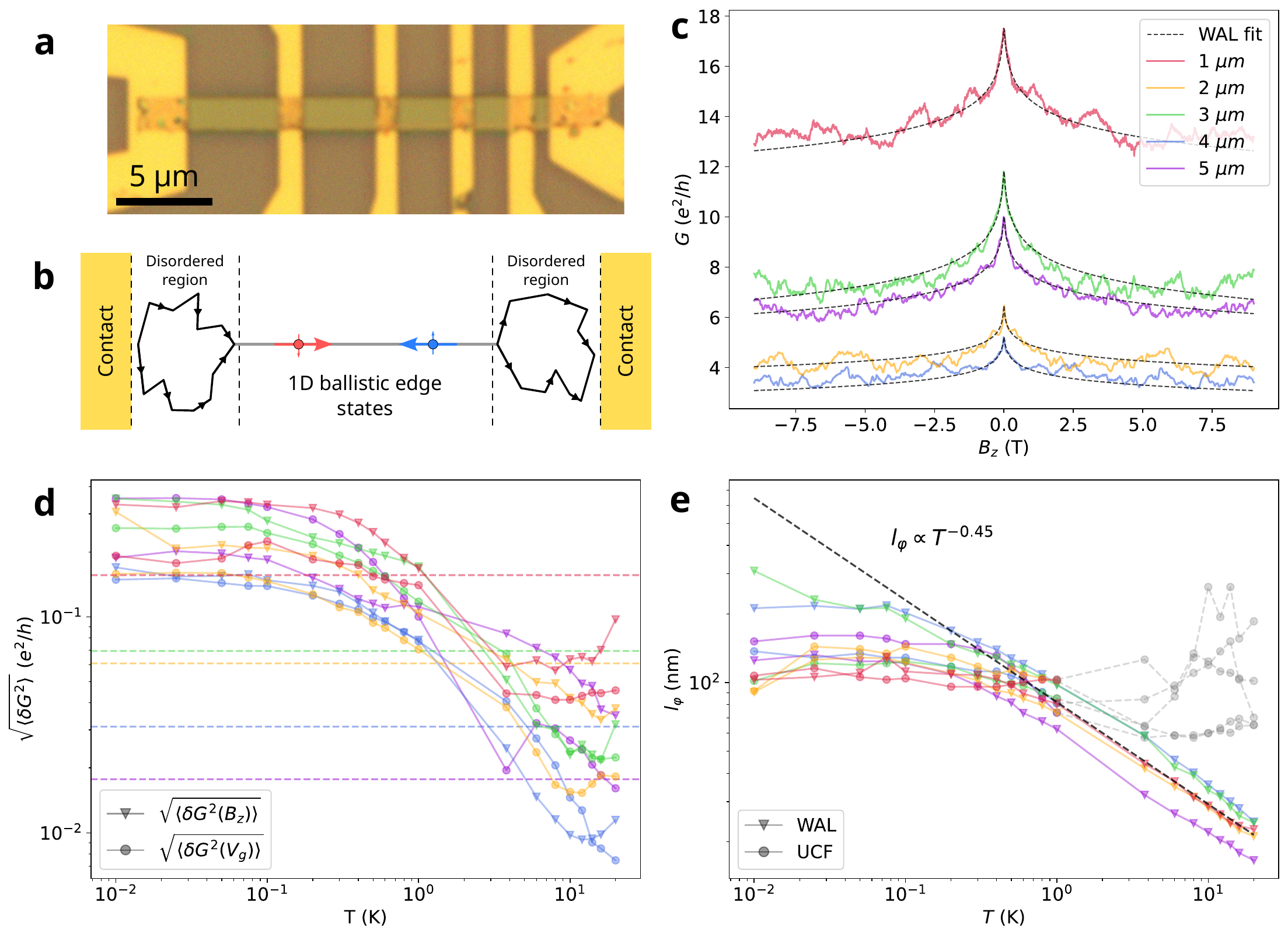}
		\caption{Coherent quantum transport measurement in sample $\mathcal{L}$. (a) Optical image of sample $\mathcal{L}$. (b) Sketch of the different sample regions contributing to the quantum transport. (c) Magnetoconductance at 10 mK of all segments for an out-of-plane magnetic field. Correspondence between the segment and curve is indicated in the legend. The dashed line is a fit of the WAL component using formula (\ref{eq:HLN}). (d) Dependence of UCF amplitudes (in units of $e^2/h$) as a function of temperature. The amplitude is deduced from the gate dependence (round symbols) or the magnetic field dependence (triangle symbols) of the UCFs. The color code is the same as (c). The horizontal dashed lines are the expected value of the UCF for a diffusive conductor of the same length and same $\alpha$ as the considered section if self-averaging is assumed. (e) $l_\varphi$ extracted from both the UCFs (round symbols) and WAL (triangle symbols), showing a power-law dependence in temperature with exponent $-0.45$. $l_\varphi$ saturates for temperatures below 100 mK. The part of the curve in gray indicates the range of temperature where UCFs cannot be reliably extracted due to experimental noise.}
		\label{fig:interf}
    \end{figure*}

	At low temperature, {\color{black}the differential resistance of all samples} exhibits a zero-bias peak, which cannot be attributed to self-heating (Sec. 7 of Supplemental Material \cite{SM}). Considering the linear regime, with a low excitation current, and for temperatures below 20 K, magnetoconductance has two main characteristics: a weak-antilocalization (WAL) peak at zero magnetic field, and sizable universal conductance fluctuations (UCFs) [Fig. \ref{fig:interf}(c) for sample $\mathcal{L}$].
	
	The low-field WAL peaks are fitted using the Hikami-Larkin-Nagaoka relation : 
	\begin{equation}
		\delta G(B) = \frac{\alpha e^2}{2\pi h}\left[ \Psi\left(\frac{1}{2} + \frac{\hbar}{4eB l_\varphi^2}\right) - \ln \left(\frac{\hbar}{4eB l_\varphi^2}\right) \right]
		\label{eq:HLN}
	\end{equation}
	with $\Psi$ the digamma function, $B$ the magnetic field, $l_\varphi$ the {\color{black}phase coherence} length and  $\alpha$ a prefactor corresponding to the number of independent channels {\color{black}in parallel} \cite{Hikami1980,Montambaux2007}. Equation (\ref{eq:HLN}) is valid for a 2D diffusive conductor in the limit of very strong spin-orbit coupling (spin-orbit length $l_{\mathrm{SO}} \ll l_\varphi$). The fitting is done in the magnetic field range {\color{black}[-1 T, 1 T] for temperatures under 2 K, and [-3 T, 3 T]} otherwise with two adjustable parameters : the prefactor $\alpha$, which is kept constant for a given segment at all temperatures, and the phase coherence length $l_\varphi$. Formula (\ref{eq:HLN}) gives an excellent agreement over the entire field range. This seems to indicate the absence of other contributions to the magnetoconductance. However{\color{black},} fits using the WAL expression in the case of quasi-1D conductors yielded both a poor agreement with the experimental data and inconsistent values for $l_\varphi$ (Sec. 6 of Supplemental Material \cite{SM}). We find relatively short $l_\varphi$ across all segments, between 100 nm and 200 nm at low temperatures [Fig. \ref{fig:interf}(e)] and values for $\alpha$ between 2 and 8, which suggests the existence of several uncorrelated electron paths in parallel \cite{Montambaux2007,Steinberg2011,Chen2011}.
	
	We now turn to the analysis of the UCFs, which were measured as a function of magnetic field [Fig. \ref{fig:interf}(c)] and gate voltage (Sec. 3 of Supplemental Material \cite{SM}). The UCFs exhibit a typical magnetic field scale, given by the half-width of the autocorrelation, which provides another determination of $l_\varphi$ \cite{Lee1987,Beenakker1988,Mohanty2003}. As shown in Fig. \ref{fig:interf}(e), these values are close to those extracted from WAL, except at temperatures above 1 K, for which the experimental noise outweighs the {\color{black}UCF} signal. 
	For all sections of sample $\mathcal{L}$, $l_\varphi$ extracted from both WAL and UCFs follows a power-law dependence down to approximately 100 mK, below which it saturates. We find an exponent of $-0.45$, consistent with the expected $-1/2$ for dephasing in 2D dominated by electron-electron interactions \cite{Montambaux2007}.\\
	\begin{figure}[tb]
		\centering
		\includegraphics[width=0.95\columnwidth]{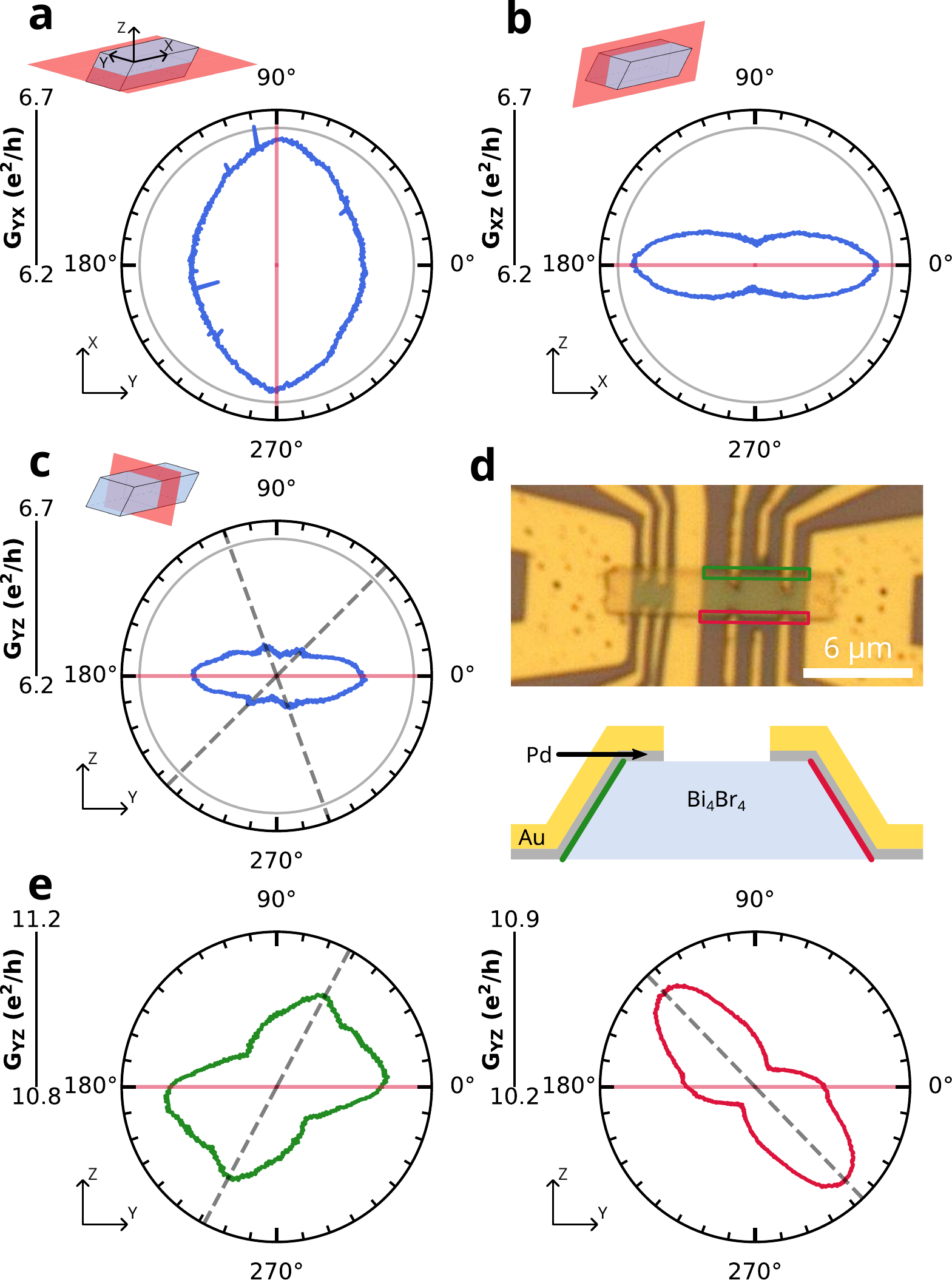}
		\caption{(a)-(c) : Magnetoconductance anisotropy at 3.8 K for the 4-$\mu$m segment of sample $\mathcal{L}$.
		(a) Top : schematic image of the flake and the $XY$ plane in which the magnetic field is applied. Bottom : variation of the conductance by sweeping the magnetic field angle in the $XY$ plane, for a magnetic field modulus of 1 T. The angle 0° corresponds to a field aligned along the $Y$ direction. Magnetoconductance is highest in the Y direction, defining an extended lobe in the perpendicular $X$ direction, as indicated by the red line. (b) Similar data for a magnetic field in the $XZ$ plane, showing a maximum magnetoconductance along the $Z$ direction. (c) Similar data for a magnetic field in the $Y$ plane, showing a maximum magnetoconductance along the $Z$ direction and side lobes, indicated by gray dashed lines, attributed to the geometry of the contacts (see the text). (d) Top : optical image of the sample $\mathcal{H}_2$. Bottom : sketch of the cross section of the flake $\mathcal{H}_2$. (e) Magnetoconductance with a magnetic field of 1 T in the $YZ$ plane for lateral contacts indicated in (d) in green (left part) or red (right part). The direction of maximum magnetoconductance is not related to the crystalline orientation but rather to the geometry of the lateral contacts.}
		\label{fig:anisotropy}
    \end{figure}
    
	Despite $l_\varphi$ being much shorter than the segment length, we find sizable fluctuation amplitudes with no systematic decrease with increasing segment length. We recall that UCF in diffusive conductors should decrease as the segment length extends beyond $l_\varphi$, due to self-averaging. By contrast{\color{black},} the absence of a systematic decrease in fluctuations amplitude with length [Fig. \ref{fig:interf}(d)] shows that the quantum interference we measure does not originate from a diffusive path within the flake. 
	
	Rather, we propose that diffuse transport occurs only in the damaged region near the contacts, roughly 100 nm in size as discussed above. Those diffusive regions are connected by 1D ballistic channels, up to 5 $\mu$m long, so that coherent transport occurs over up to 5 $\mu$m. We conjecture that it is the transmission from contacts to ballistic channel that is modulated by quantum interference in the disordered regions. This is what leads to the absence of self-averaging and to the high field scale of UCF and WAL. This scenario was already pointed out in the context of the conductance of edge states in the integer quantum Hall effect \cite{Buttiker1988} and is qualitatively reproduced by our numerical simulations [Fig. \ref{fig:Simul}(a) and section 5 in Supplemental Material \cite{SM}).
	
	To confirm this picture, we studied the dependence of quantum interference on magnetic field orientation. We performed magnetoconductance measurements on samples $\mathcal{L}$ using a vector magnet, allowing it to apply magnetic fields in {\color{black}arbitrary} directions up to 1 T. We present field cuts in three different planes : in-plane, longitudinal and transverse, all taken at 3.8 K to remove the contribution from conductance fluctuations, which complicates the analysis [Fig. \ref{fig:anisotropy}(a)-\ref{fig:anisotropy}(c)]. The first two cuts clearly highlight a near absence of magnetoconductance along the $b$ axis of the flake [$X$ direction of Figs. \ref{fig:anisotropy}(a) and  \ref{fig:anisotropy}(b)]. The longitudinal and transverse cuts reveal that the decrease of conductance with magnetic field is maximum for an out-of-plane field, leading an extended lobe in the polar plot in the perpendicular direction (indicated by the {\color{black}red} lines in Fig. \ref{fig:anisotropy}), demonstrating that the surface of diffusion essentially lies in the plane of the sample. This behavior, together with the absence of self-averaging of the UCFs, rules out diffusion on the sides surfaces of Bi$_4$Br$_4$ as the source of the observed quantum interferences.

	In addition to the large out-of-plane magnetoconductance observed in the transverse cut, several devices contacted from the top exhibit other lobes [gray dashed lines in Figs. \ref{fig:anisotropy}(c), \ref{fig:anisotropy}(f) and \ref{fig:anisotropy}(g)] that are not related to {\color{black}any} specific crystalline direction of Bi$_4$Br$_4$. {\color{black}Rather, these} axes correspond to {\color{black}the planes} of the electrodes deposited on the lateral side of the flake. Indeed, the disordered region induced by the contacts follows the slope of the etched crystal, which defines the plane of diffusion for the WAL [Fig. \ref{fig:anisotropy}(e)]. Moreover, for sample $\mathcal{H}_2$, with a Hall bar geometry, we find that, for contacts on the side of the flakes, only one of the side lobes is present and that it is more prominent than the out-of-plane one. This confirms that each side lobe comes from one flank of the flake. {\color{black}In contrast, samples} contacted from below showed only an out-of-plane magnetoconductance, confirming that the origin of the side lobes lies in the contact method.
	
	In conclusion{\color{black},} magnetotransport at low temperature and its anisotropy with magnetic field orientation confirms that the regions determining the mesoscopic fluctuations are the diffusive contact regions, leading to a field-dependent transmission to 1D ballistic channels. 

%%%%%%%%%%%%%%%%%%% Negative 4W resistance %%%%%%%%%%%%%%%%%%%%%%%%%%%%%%%%%%%%%%%%
	\section{Negative four-wire resistance}
	
    While the behavior of samples contacted from above and from below are similar on many aspects, there are a few clear differences. The first is the presence of a higher interface resistance for bottom contacts, and the second is the absence of a four-wire negative differential resistance in top-contacted samples.
    \begin{figure}[tb]
		\centering
		\includegraphics[width=\columnwidth]{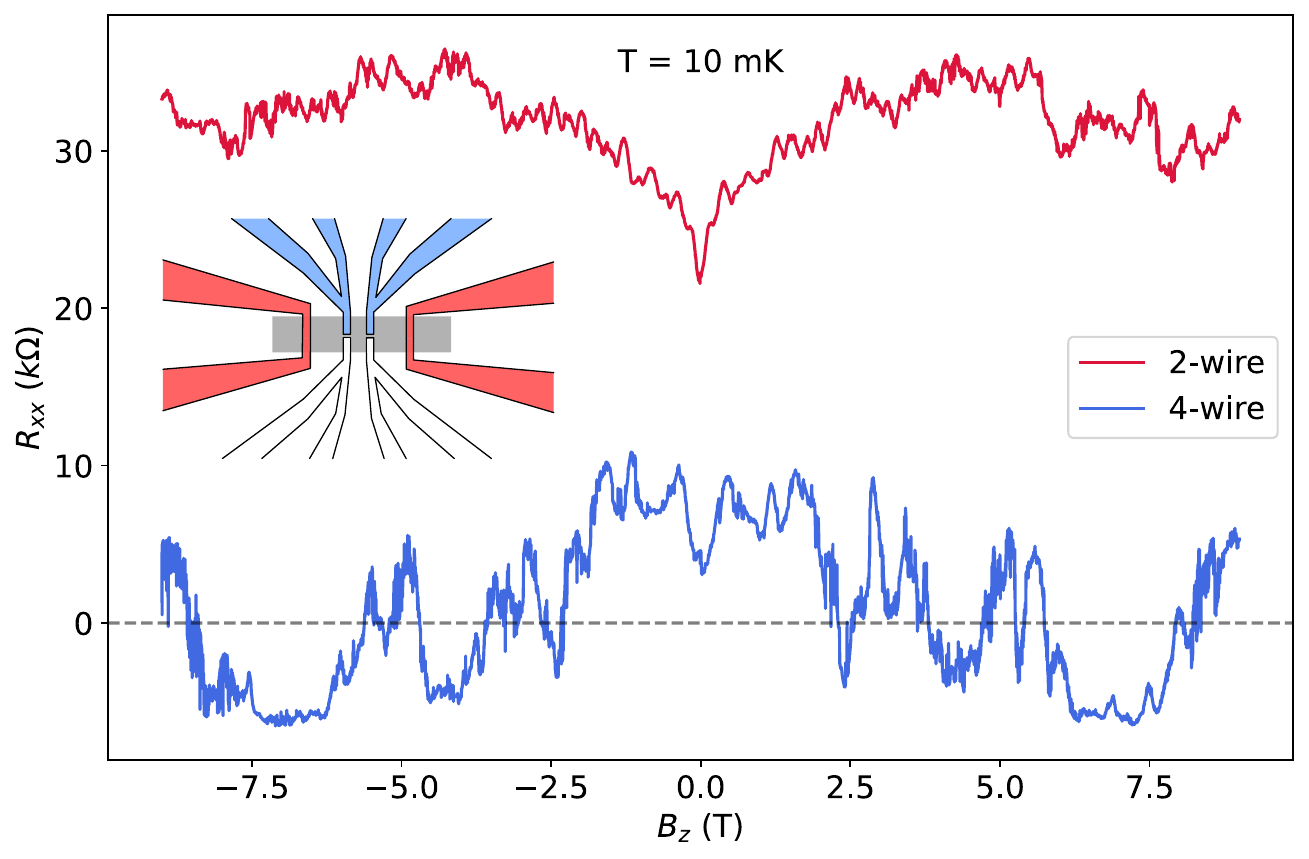}
		\caption{Two- and four-wire {\color{black}longitudinal} resistances of sample $\mathcal{H}_1$ as a function of magnetic field at $T = 10$ mK, measured with a 10 nA {\color{black}ac} current. The four-wire measurement is negative in {\color{black}several field ranges}. Inset : sample geometry with the Bi$_4$Br$_4$ flake in gray, current injection in red and voltage probes in blue. For the two-wire measurement, the voltage {\color{black}probes} are the red wires.}
		\label{fig:Neg_R}
	\end{figure}
	
    Indeed, we observe {\color{black}negative four-wire resistance} fluctuations that extend between -6 and 10 k$\Omega$ with a two-wire resistance of 20-35 k$\Omega$ {\color{black}for} $\mathcal{H}_1$ (Fig. \ref{fig:Neg_R}). This corresponds to a {\color{black}sizable} fraction of the theoretical limit of $\pm R_{2 \mathrm{wires}}$, predicted for ballistic conduction measured with moderately invasive voltage probes \cite{Buttiker1988a}. Indeed, the two-wire resistance {\color{black}measured through the voltage probes indicated in blue in Fig. \ref{fig:Neg_R}} is high {\color{black}(approximately 150 k$\Omega$)}, {\color{black}which may indicate a} weakly perturbative {\color{black}contact to} the edge states that can favor the appearance of these negative fluctuations. 
    {\color{black}Indeed,} contacts from below may allow for the hinge states to be preserved to a certain extent, if they are not cut by a large Pd/Bi alloy region (Fig. \ref{fig:EDX}).
    We emphasize that this prediction of a four-wire negative differential resistance for ballistic systems has not been reported by many studies. Experiments on single-walled carbon nanotubes have found negative fluctuations reaching the theoretical limit \cite{Gao2005}, as well as measurements on high-mobility GaAs 2DEGs, which were observed only in a specific field range \cite{Timp1987,dePicciotto2001}. Both of these experiments featured exceptionally clean systems with noninvasive voltage probes, which contrasts with the disorder induced at the contacts with Bi$_4$Br$_4$. Therefore, this observation of negative differential resistance in our samples could be a consequence of the expected topological protection of hinge states {\color{black}against} elastic disorder.

%%%%%%%%%%%%%%%%%%% AB effect %%%%%%%%%%%%%%%%%%%%%%%%%%%%%%%%%%%%%%%%%%%%%	
	\section{Aharonov-Bohm interference}
	
	In addition to the aperiodic UCFs and WAL displayed in the magnetoconductance, several samples also display periodic oscillations, leading to well-defined peaks in their Fourier transform (FFT).
	\begin{figure}[tb]
		\centering
		\includegraphics[width=\columnwidth]{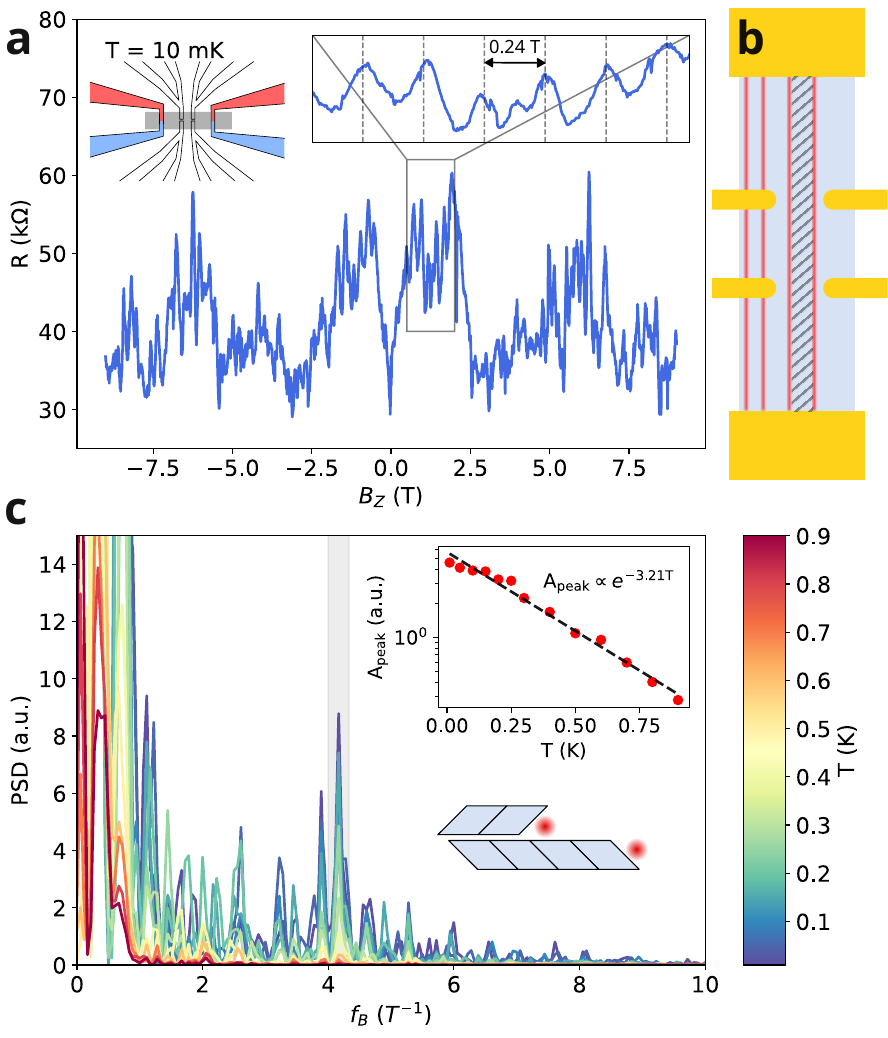}
		\caption{Aharonov-Bohm effect of sample $\mathcal{H}_1$. (a) Magnetoconductance at low temperature exhibiting periodic conductance oscillations, with a field period of 0.24 T. Left inset : configuration of the contacts used for current injection (red) and voltage probing (blue). Right inset : enlargement of a reduced field range. (b) Schematic picture of the Bi$_4$Br$_4$ flake (blue) with its contacts (yellow). The red lines represent 1D states, some of which are interrupted by lateral contacts. The two central 1D states delimit a surface (hatched area) defining the magnetic field periodicity of the AB effect. (c) Power spectral density of the magnetoconductance of sample $\mathcal{H}_1$ at different temperatures. The gray area indicates the peak corresponding to the observed conductance {\color{black}oscillations}, taking into account an uncertainty of $\pm$ $20 \times 20$ nm$^2$. Top inset : temperature dependence of the FFT peak area. It is consistent with an exponential decay with temperature, of exponent $-3.21$. Bottom inset : schematic image of a cross section of Bi$_4$Br$_4$ exhibiting {\color{black}a configuration of edge states} compatible with the observed AB periodicity. The blue diamonds represent individual crystalline unit cells and the red dots show the position of edge states.}
		\label{fig:AB_H1}
	\end{figure}
	
	\begin{figure}[tb]
		\centering
		\includegraphics[width=\columnwidth]{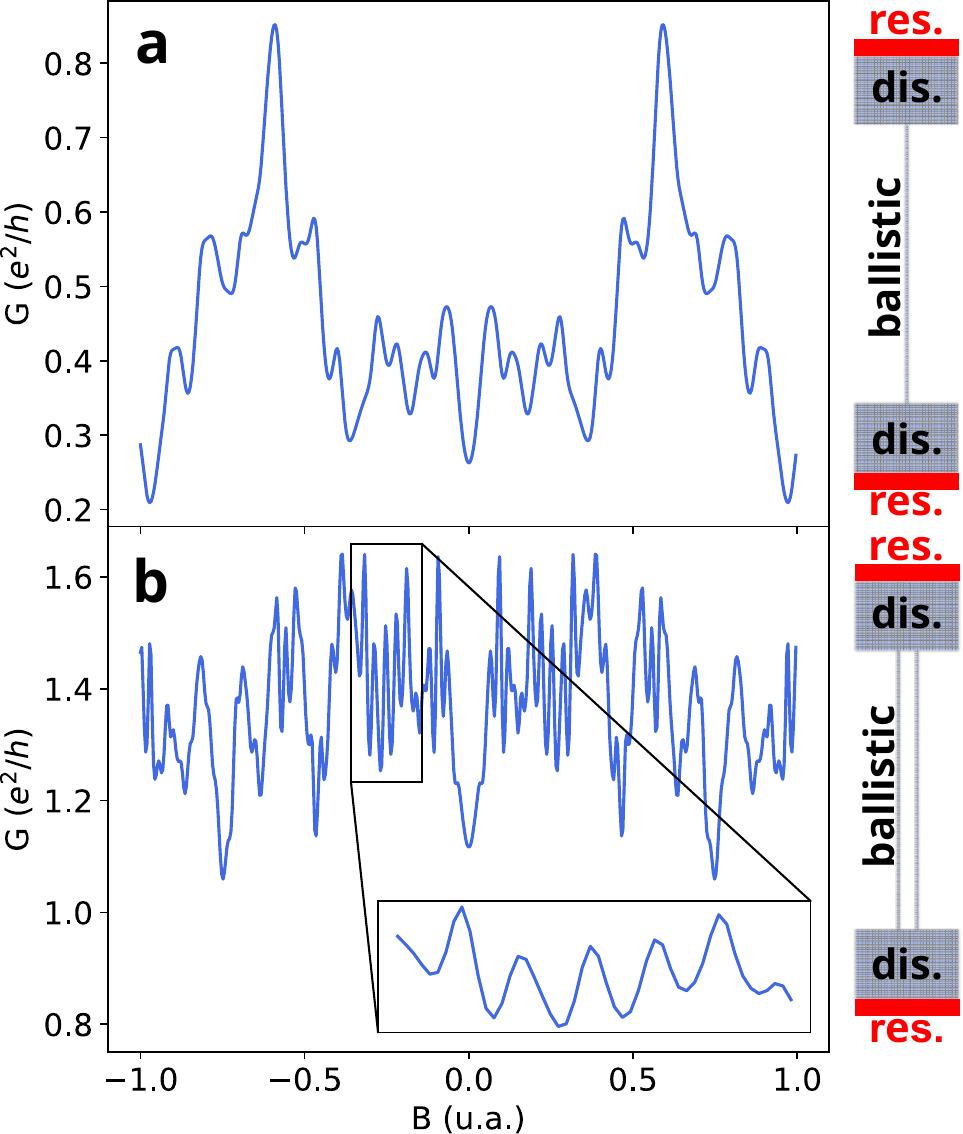}
		\caption{Numerical simulations of the conductance of ballistic channels connected to reservoirs via a disordered coherent region realized with the KWANT software package \cite{Groth2014}. (a) Case of a single ballistic channel showing WAL and UCFs. Right : tight binding network used to calculate the conductance (Sec. 5 of Supplemental Material \cite{SM}). The long, narrow region is the ballistic channel (noted ballistic). It is terminated by two diffusive coherent regions (noted dis.). The reservoirs are represented by the red rectangles (noted res.).  b) The same for two ballistic channels in parallel connected to reservoirs via a disordered coherent region. The conductance exhibits AB oscillations on top of WAL and UCFs. Right : corresponding tight binding network \cite{SM}.}
		\label{fig:Simul}
	\end{figure}
	
	In particular the longitudinal magnetoresistance of sample $\mathcal{H}_1$ shows a very clear periodicity of around 0.24 T at low temperatures leading to a FFT peak at 4 T$^{-1}$ [Fig.\ref{fig:AB_H1}(a) and \ref{fig:AB_H1}(c)]. {\color{black} Note that the periodicity of the oscillations does not depend on temperature but the amplitude does (Sec. 4 of Supplemental Material \cite{SM}).} We interpret this as Aharonov-Bohm (AB) {\color{black}interference} \cite{Washburn1986} between two 1D states of length 6 $\mu$m and {\color{black}separated} by 2.7 nm, allowed by the partially coherent contact region, as confirmed by numerical simulations [Fig. \ref{fig:Simul}(b) and {\color{black} Sec. 5 of Supplemental Material} \cite{SM}). One possible reason for this particularly clear signature of AB interference in this sample is the small separation between the lateral electrodes, which selects only a few channels that can carry current coherently across the flake without being interrupted by a contact [Fig.\ref{fig:AB_H1}(b)]. The good visibility of the AB oscillations is also favored by the small lateral distance between edge states, comparable to the mean free path in the diffusive contact [Sec. 5 of Supplemental Material \cite{SM}]. {\color{black}Besides, we} find an exponential decay with temperature of the AB amplitude [inset in Fig. \ref{fig:AB_H1}(c)). This temperature behavior has been reported in quasi-1D ballistic rings \cite{Hansen2001}, 3D topological insulators \cite{Dufouleur2013,Jauregui2016,Behner2023} and hinge state loops in thin MoTe$_2$ flakes \cite{Pan2023}. This dependence is, however, at odds with the result of Hossain \textit{et al.} \cite{Hossain2024} on thin flakes of Bi$_4$Br$_4$, which instead exhibits a power-law dependence that is attributed to the extremely long, {\color{black}effectively} temperature-independent, phase coherence length of edge states. The functional difference between the decay of the UCFs and AB oscillations with temperature in our samples suggests that their coherence is limited by distinct mechanisms. One would expect the amplitude of AB oscillations to be proportional to the probability of an electron trajectory to complete a loop without phase-breaking event $\exp(- P/l_\varphi)$, with $P$ the perimeter of the closed trajectory \cite{Washburn1986}. The exponential decay of the peak amplitude in the FFT with temperature, therefore, implies a $1/T$ dependence of the phase coherence length and taking $P=12\ \mu$m, $l_\varphi=3.7\ \mu$m at 1 K. The fact that we can measure {\color{black}such a} long phase coherence length implies that phase breaking occurs predominantly within the hinge states, so that the electron trajectories contributing the most to the AB effect do not explore the diffusive regions much, thanks to the very small separation between hinge states (2.7 nm). Our results emphasize the coherence of electronic transport in relatively long Bi$_4$Br$_4$ flakes.
    \begin{figure}[tb]
		\centering
		\includegraphics[width=\columnwidth]{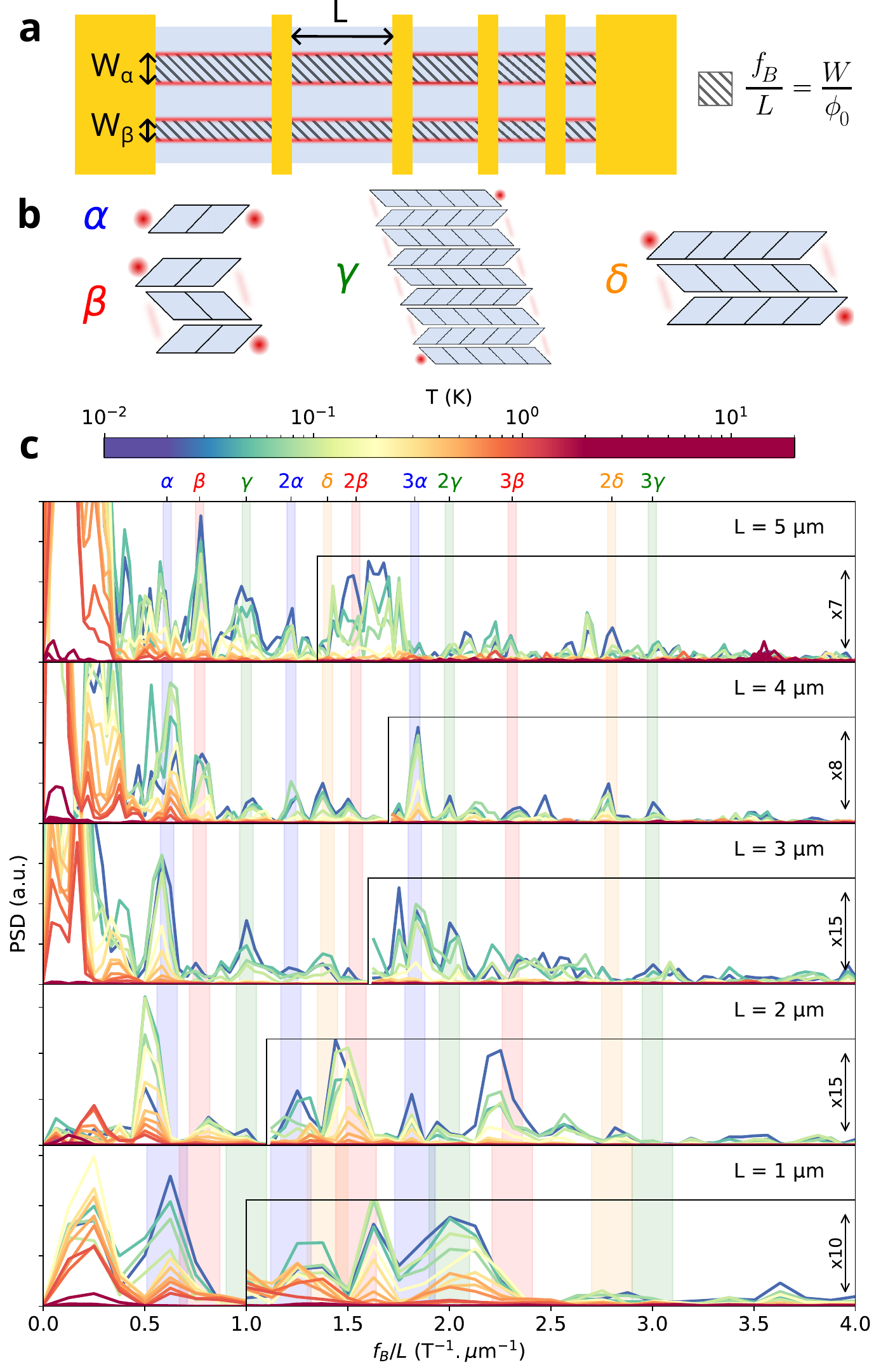}
		\caption{AB effect in sample $\mathcal{L}$. (a) Schematic picture of the sample with the Bi$_4$Br$_4$ flake in blue {\color{black}and} the contacts in yellow. The red lines represent 1D states, which are separated by a distance $W$ and interrupted by the contacts. For a segment of length $L$, the AB magnetic field period is $\Phi_0/(W L)$, thus leading to a magnetic frequency $f_B=W L / \Phi_0$. (b) {\color{black}Hinge state configurations consistent with the observed AB peaks in the FFT [see (c)], where a blue diamond represents a crystalline unit cell and the red dots correspond to the hinge states.} (c) Power spectral density of the magnetoconductance of the different segments of sample $\mathcal{L}$ and for temperatures between 10 mK and 10 K, indicated by the color bar. The FFT frequencies are rescaled by the length of the corresponding segment. Shaded areas correspond to an uncertainty of $\pm$ $20 \times 20$ nm$^2$ around a few identified peaks [see (b) and Sec. 4 of Supplemental Material \cite{SM}], {\color{black}denoted $\alpha$ (blue), $\beta$ (red), $\gamma$ (green) and $\delta$ (orange), and their harmonics}.}
		\label{fig:AB_L}
	\end{figure}
	
	Strikingly, for device $\mathcal{L}$, several of the FFT peaks correspond to a field scale proportional to the length of the considered segment. Hence{\color{black},} if the magnetic frequency is rescaled by the segment length [Fig. \ref{fig:AB_L}(b)], it is possible to identify some peaks in the FFT of different segments that correspond to the same rescaled magnetic frequency. We interpret this as AB {\color{black}interference} around surfaces of fixed widths across each segment. We postulate that these {\color{black}oscillations} originate from coherent 1D hinge states that extend over the entire length of the flake, {\color{black}which were subsequently divided into several segments by the fabricated contacts}. Consequently, from segment to segment, there is a constant lateral separation between the hinge states, as illustrated in Fig. \ref{fig:AB_L}(a). {\color{black}In fact, the magnetic frequencies identified in the FFT spectrum can be related to some specific configurations of hinge states that are consistent with the crystalline structure of Bi$_4$Br$_4$ [Fig. \ref{fig:AB_L}(b)]}. As the AB effect is visible for all segments, including the longest ones, the phase coherence length of the edge states must be of the order of 5 $\mu$m or more. However, the limited phase coherence in the {\color{black}contact regions} induces an upper cutoff to the lateral distance between 1D channels allowing for AB interferences, such that we do not observe peaks in the Fourier spectrum corresponding to distances greater than a few tens of nanometers. Moreover, we attribute the absence of a clear magnetic field periodicity in the magnetoconductance signal to the multiplicity of frequencies present in the Fourier transform, which are due to the many 1D coherent channels of the flake that may interfere with each other. The temperature dependence of the identified peaks exhibits an exponential decay consistent with a $1/T$ dependence of phase coherence length and a value at 1 K similar to the one of sample $\mathcal{H}_1$. Moreover the existence of AB interference in all segments of $\mathcal{L}$ up to 5 µm is consistent with long phase coherence length.

	\section{Conclusion}
    	We have probed the electronic transport in micrometer-size Bi$_4$Br$_4$ flakes and have demonstrated a strongly anisotropic transport. We have highlighted two distinct types of contributions to low-temperature coherent transport and argue that they {\color{black}reveal the existence} of long coherent 1D hinge states contacted to reservoirs via a small coherent disordered region. The first type of {\color{black}contribution} yields universal conductance fluctuations and weak antilocalization, {\color{black}arising} from phase coherence in the region close to the contacts, {\color{black}which} modulates transmission through the 1D states. The second type of {\color{black}contribution} is the periodic Aharonov-Bohm conductance oscillations due to interference between adjacent 1D coherent channels. We emphasize that both of these effects are due to the very nature of the contacts, which exhibit some phase coherence despite being rather invasive. {\color{black}These} effects indicate that {\color{black}the} 1D states are ballistic and coherent over several micrometers at low temperatures. The signature of the 1D ballistic states, located at step edges distributed at the surface of the Bi$_4$Br$_4$ crystal, and their interference are thus paradoxically revealed thanks to the coherent diffusive nature of the contact regions. {\color{black}It would be interesting to achieve the opposite limit of clean, crystal-preserving contacts in small samples with a reduced number of edge states, in order to access the conductance quantization regime}. {\color{black}Also, the} signature of the helical nature of these 1D states in multilayer samples remains to be demonstrated. {\color{black}Lastly}, the use of superconducting contacts to enter the regime of proximity-induced superconductivity is of great interest, with expected perfect Andreev reflection, and would further enhance the potential of Bi$_4$Br$_4$ as a second-order topological insulator. 
	
	\begin{acknowledgements}
         The authors acknowledge B. Wieder, E. Gerber, T. Wakamura, M. Kociak, M. Aprili and J. Esteve for fruitful discussions, L. Bugaud's involvement in the initial steps of the project, and technical help from S. Autier-Laurent and R. Weil. This work was funded by JSPS KAKENHI (Grants No. 21H04652, No. 21K18181 and No. 21H05236), JST CREST (Grant No. JPMJCR20B4), the French Renatech network, the CPER Hauts de France project IMITECH, the Métropole Européenne de Lille, and the European Union through the BALLISTOP ERC 66566 advanced grant. 
    \end{acknowledgements}
    
    \section*{DATA AVAILABILITY}
    The data that support the findings of this article are openly available \cite{zenodo}.


\begin{thebibliography}{}
\bibitem{reciprocalCond} This method of finding the conductance matrix can only be applied if describing the sample as a network of resistors is adequate, meaning that the transport between nodes is reciprocal.
\end{thebibliography}
\end{document}

\title{Supplemental material : Quantum Coherent Transport of 1D Ballistic States in Second Order Topological Insulator Bi$_4$Br$_4$}

    \author{Jules Lefeuvre}
    \email{jules.lefeuvre@universite-paris-saclay.fr}
    \affiliation{Université Paris-Saclay, CNRS, Laboratoire de Physique des Solides, 91405, Orsay, France.}
    \author{Masaru Kobayashi}
    \affiliation{Materials and Structures Laboratory, Institute of Science Tokyo, Yokohama, Japan.}
    \author{Gilles Patriarche}
    \author{Nathaniel Findling}
    \affiliation{Université Paris-Saclay, CNRS, Centre de Nanosciences et de Nanotechnologies, 91120, Palaiseau, France.}
    \author{David Troadec}
    \affiliation{Université de Lille, CNRS, Institut d'Electronique, de Microélectronique et de Nanotechnologie,  59652, Villeneuve d'Ascq, France}
    \author{Meydi Ferrier}
    \author{Sophie Guéron}
    \author{Hélène Bouchiat}
    \affiliation{Université Paris-Saclay, CNRS, Laboratoire de Physique des Solides, 91405, Orsay, France.}
    \author{Takao Sasagawa}
    \affiliation{Materials and Structures Laboratory, Institute of Science Tokyo, Yokohama, Japan.}
    \author{Richard Deblock}
    \email{richard.deblock@universite-paris-saclay.fr}
    \affiliation{Université Paris-Saclay, CNRS, Laboratoire de Physique des Solides, 91405, Orsay, France.}
% \date{}

\begin{abstract}
	This part provides supplementary information for the article "Quantum Coherent Transport of 1D ballistic states in second order topological insulator Bi$_4$Br$_4$". The following topics are discussed: the crystalline quality of Bi$_4$Br$_4$ flakes, resistance networks of two other samples at different temperatures, the length dependence of resistance for samples $\mathcal{L}$ and $\mathcal{L}_2$, {\color{black}extended data and analysis on the temperature dependence of the Aharonov-Bohm effect}, details of the numerical simulations presented in the main text, the fitting of the magnetoconductance using the formula for 1D weak anti-localization, {\color{black} a presentation of zero-bias anomaly data and comparison with theoretical predictions, and finally extended data on the absence of Hall effect and Shubnikov de Haas oscillations}.
\end{abstract}
    
    \maketitle

    \begin{figure*}[tb]
		\centering
		\includegraphics[width=0.85\textwidth]{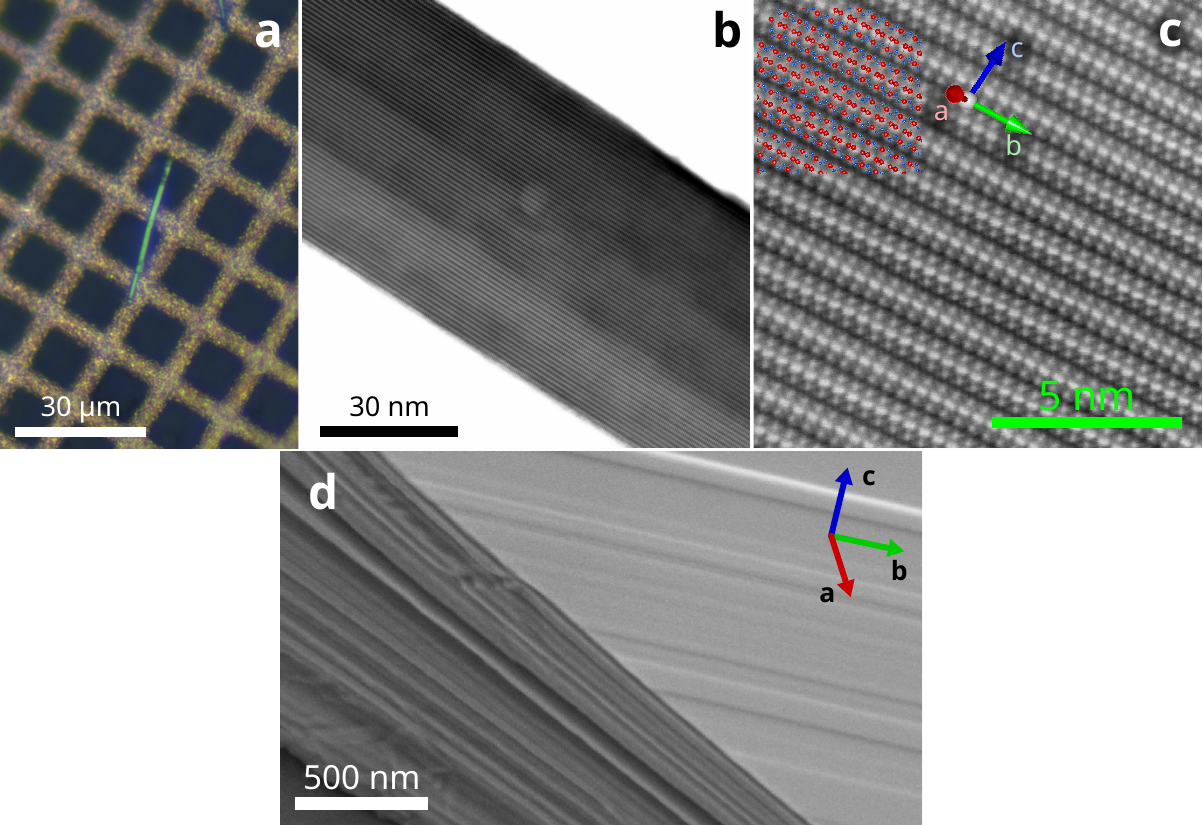}
		\caption{a) Optical picture of an exfoliated Bi$_4$Br$_4$ ribbon on a TEM grid. b) Bright Field STEM image of the flake. c) Atomic resolution HAADF-STEM image revealing alignment of the atomic columns near the a axis. On the top-left the atomic model of Bi$_4$Br$_4$ is superimposed to the STEM image. d) SEM image of a typical Bi$_4$Br$_4$ flake showing a striated top surface.}
		\label{fig:cristal_quality}
    \end{figure*}
    
    \begin{figure}[tb]
		\centering
		\includegraphics[width=\columnwidth]{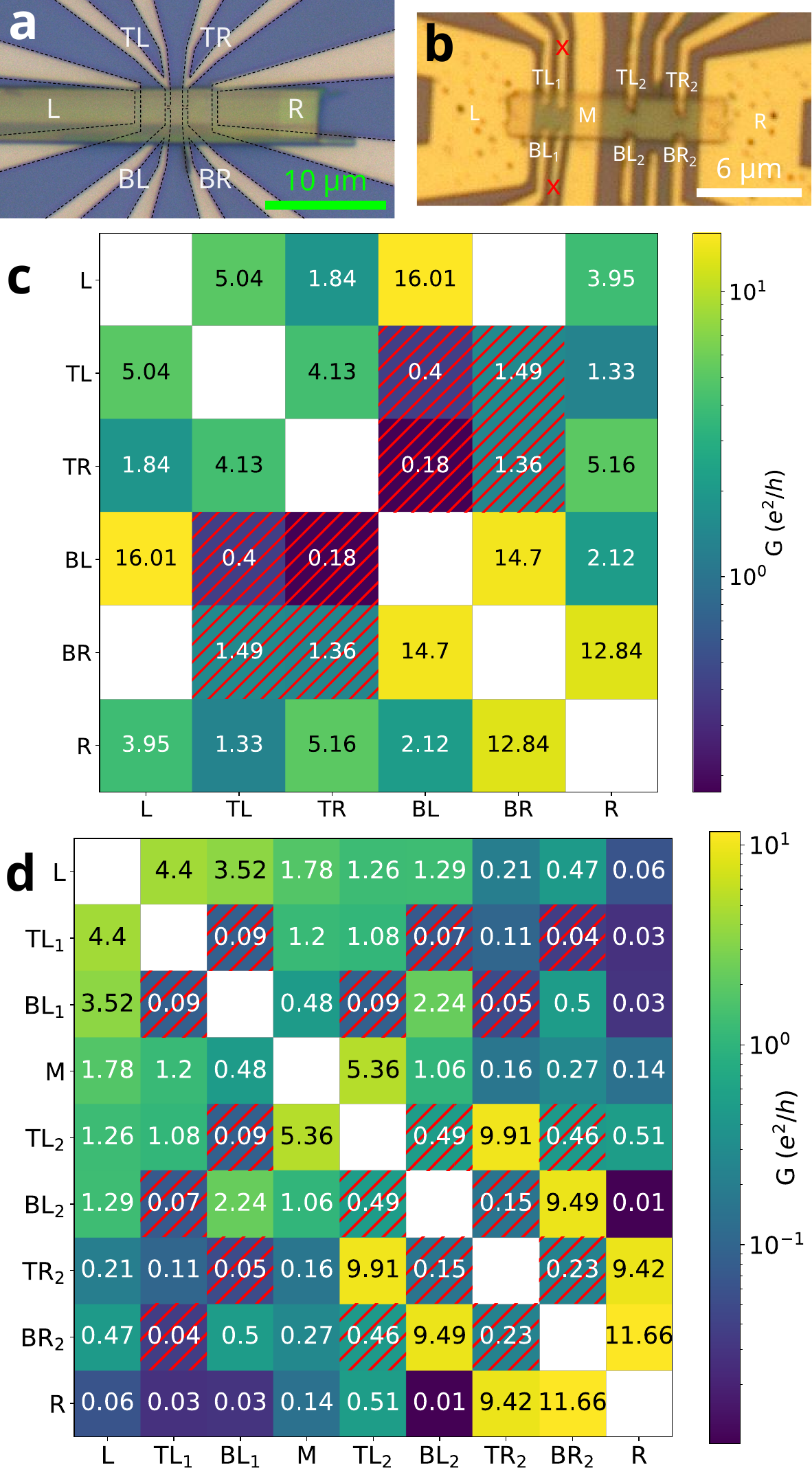}
		\caption{a) Optical image of sample $\mathcal{H}_2$. b) Optical image of sample $\mathcal{H}_3$. c) Conductance matrix of sample $\mathcal{H}_3$ taken at 300 K. d) Conductance matrix of sample $\mathcal{H}_2$ taken at 4 K. For both matrices, transverse conductances are hatched in red.}
		\label{fig:RmatH2H3}
    \end{figure}
    
    \begin{figure}[tb]
		\centering
		\includegraphics[width=\columnwidth]{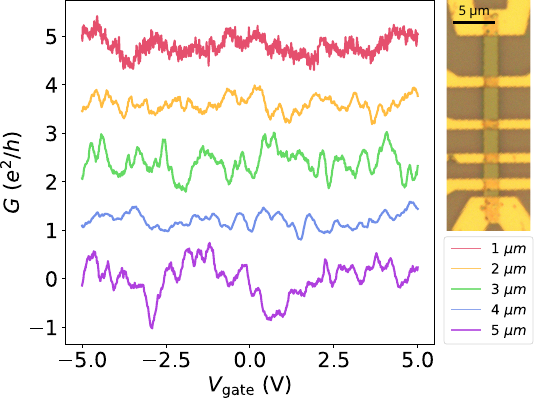}
		\caption{Gate dependence of the conductance at 10 mK for all segments of sample $\mathcal{L}$. The mean value is subtracted and the curves are shifted for clarity. Right : optical picture of sample $\mathcal{L}$.}
		\label{fig:UCF_Vg}
	\end{figure}
	
	\begin{figure}[tb]
		\centering
		\includegraphics[width=\columnwidth]{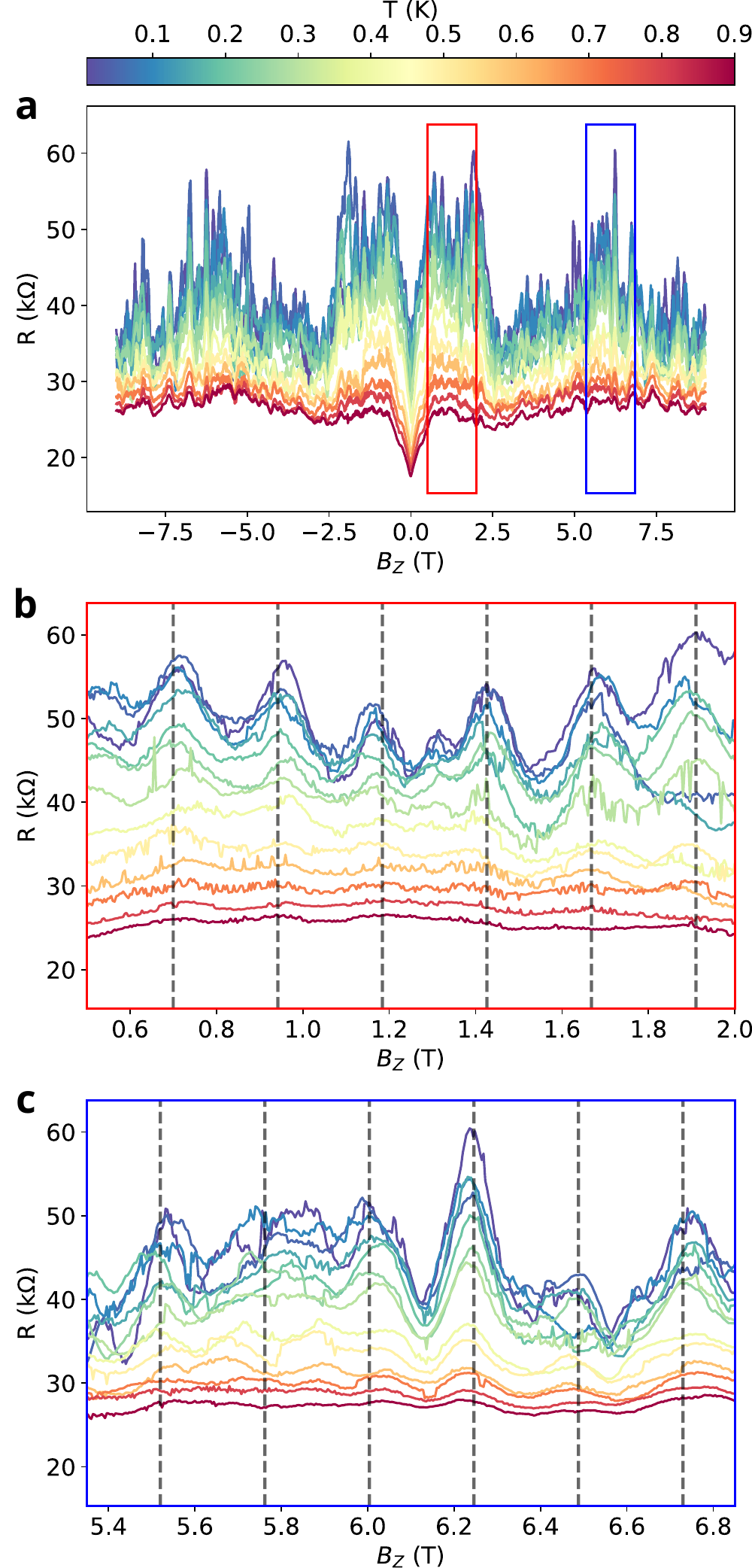}
		\caption{\color{black} a) Magnetoresistance of sample $\mathcal{H}_1$ at different temperatures from 10 mK to 900 mK, indicated by the color scale. b) and c) show two reduced magnetic ranges, indicated by the red and blue rectangles on a). The data show that the period of AB oscillations does not depend on temperature, but that their amplitude decreases rapidly with temperature.}
		\label{fig:AB_H1}
	\end{figure}
	
	\begin{figure*}[tb]
		\centering
		\includegraphics[width=0.95\textwidth]{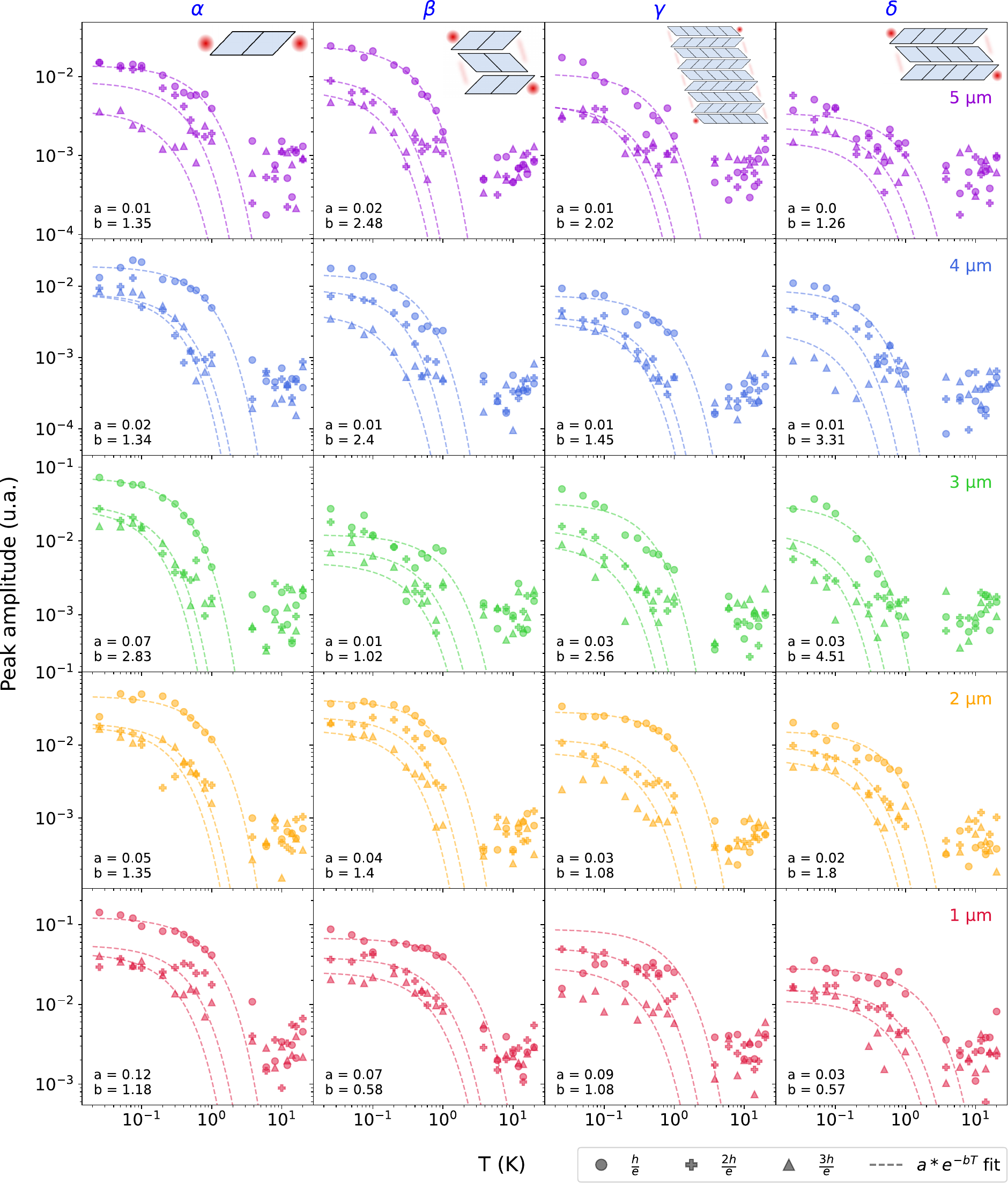}
		\caption{Temperature dependence of the peak amplitude of the Fourier transform corresponding to AB effect in the conductance fluctuations of sample $\mathcal{L}$ for all segments. For each segment and for each family of peak ($\alpha$, $\beta$, $\gamma$ and $\delta$), the first (circle), second (cross) and third (triangle) harmonics are fitted using an exponential decay. Fitting to $a e^{-bT}$ is done for the first harmonic, the n$^\mathrm{th}$ harmonics are allowed only a 5\% adjustment around $a / n$ and $b \times n$. The reference fit is done on the second harmonic of $\gamma$ for the 1 $\mu$m section. Insets : stacking picture of the cross-section of the sample surface, consistent with the considered peak, where a blue diamond represents a crystalline unit cell and the red dots the position of edge states compatible with the observed AB periodicity.}
		\label{fig:A-B_amp}
	\end{figure*}
	
	\begin{figure}[tb]
    	\centering
    	\includegraphics[width=\columnwidth]{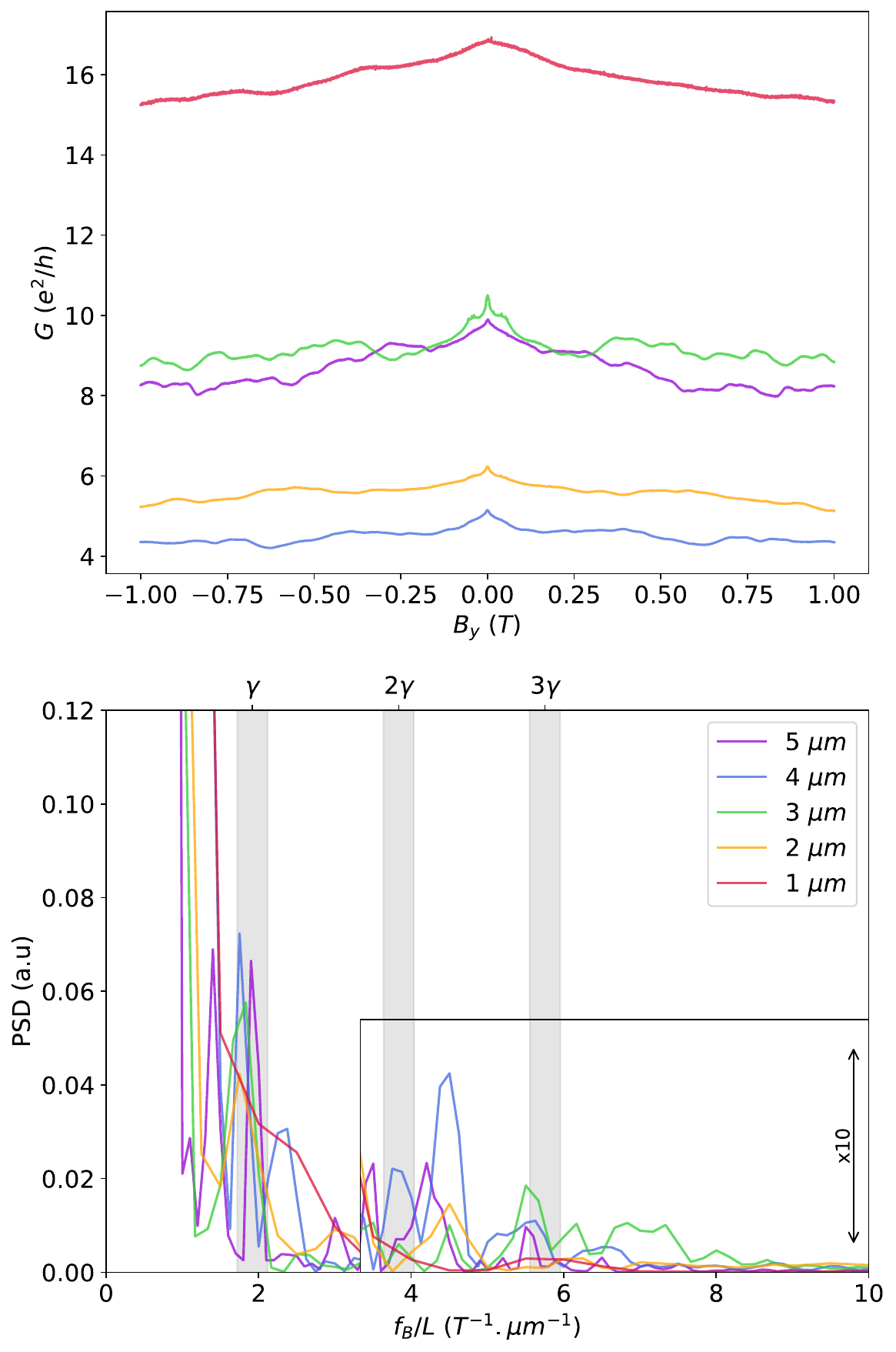}
    	\caption{Aharonov-Bohm oscillations in $\mathcal{L}$ with in-plane transverse magnetic field. \textbf{a)} Magnetoconductances of each segment at 10 mK. \textbf{b)} Power spectral density of the magnetoconductance signals with FFT frequencies rescaled by the length of each segment. Peaks corresponding to the expected in-plane response of $\gamma$ are highlighted. The PSD amplitude is multiplied by 10 at higher $f_B$ for better visibility of the peaks.}
    	\label{fig:A-B_L1_By}
    \end{figure}
	
	\begin{figure}[tb]
		\centering
		\includegraphics[width=\columnwidth]{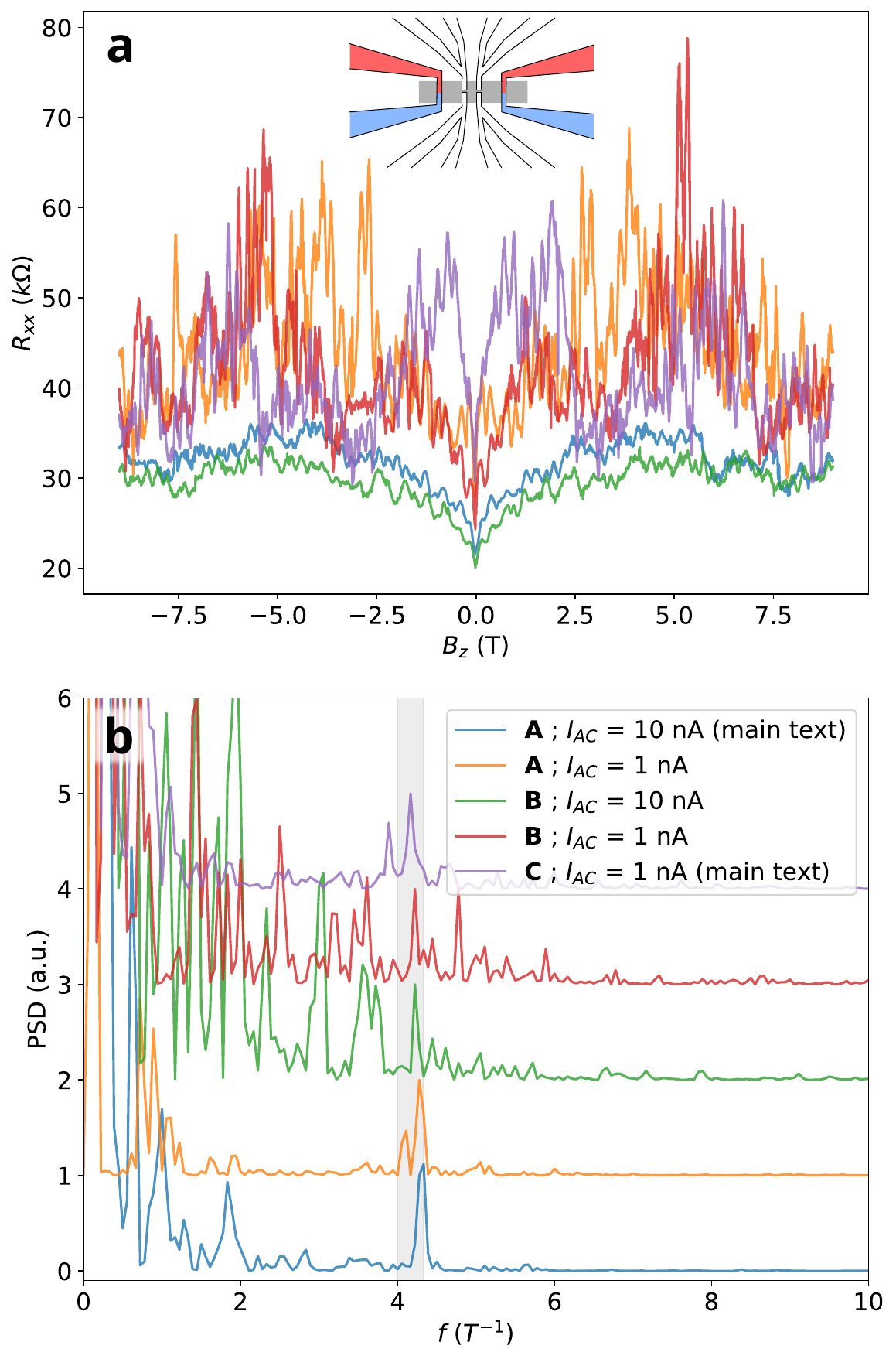}
		\caption{a) Magnetoresistance curves of sample $\mathcal{H}_1$ for different disorder configurations and ($\mathbf{A}$, $\mathbf{B}$ and $\mathbf{C}$) and current biases (1 nA or 10 nA). b) PSDs of all magnetoresistance curves, normalized by their values at $f_B=4.16$ T$^{-1}$. Despite differences in disorder realizations, all measurements show a peak in their Fourier transform corresponding to AB oscillations of period 0.24 T.}
		\label{fig:AB-Hall_disorder}
        \end{figure}
        
	\begin{figure*}[tb]
		\centering
		\includegraphics[width=0.9\textwidth]{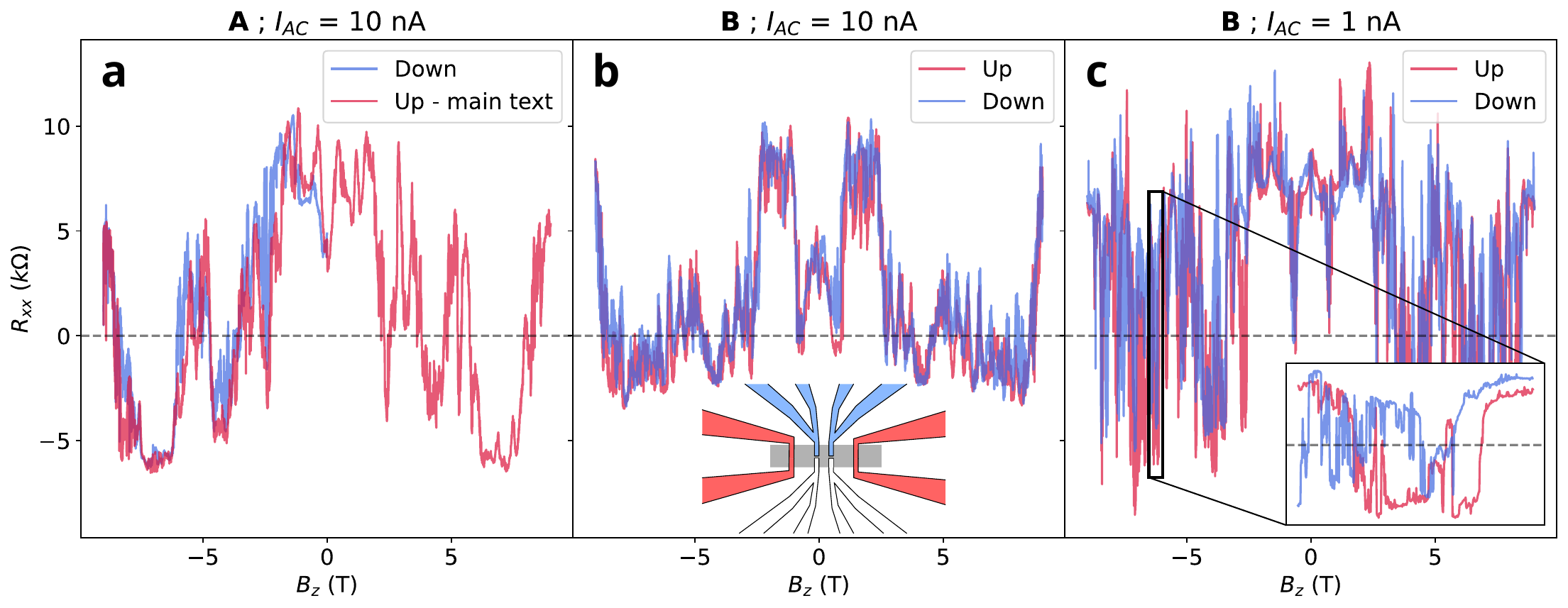}
		\caption{Reproducibility of the negative 4-wire resistance for sample $\mathcal{H}_1$. $\mathbf{A}$ (panel a)) and $\mathbf{B}$ (panels b) and c)) denote two disorder realizations, before and after a power cut. a) and b) are magnetoresistance curves taken with 10 nA AC current polarization, whereas c) was taken with 1 nA AC polarization. While a) and b) show decent repeatability between up and down field sweeps, c) shows significant telegraphic noise in certain field regions.}
		\label{fig:neg-dVdI_fluct}
    \end{figure*}
    
    \begin{figure}[tb]
		\centering
		\includegraphics[width=\columnwidth]{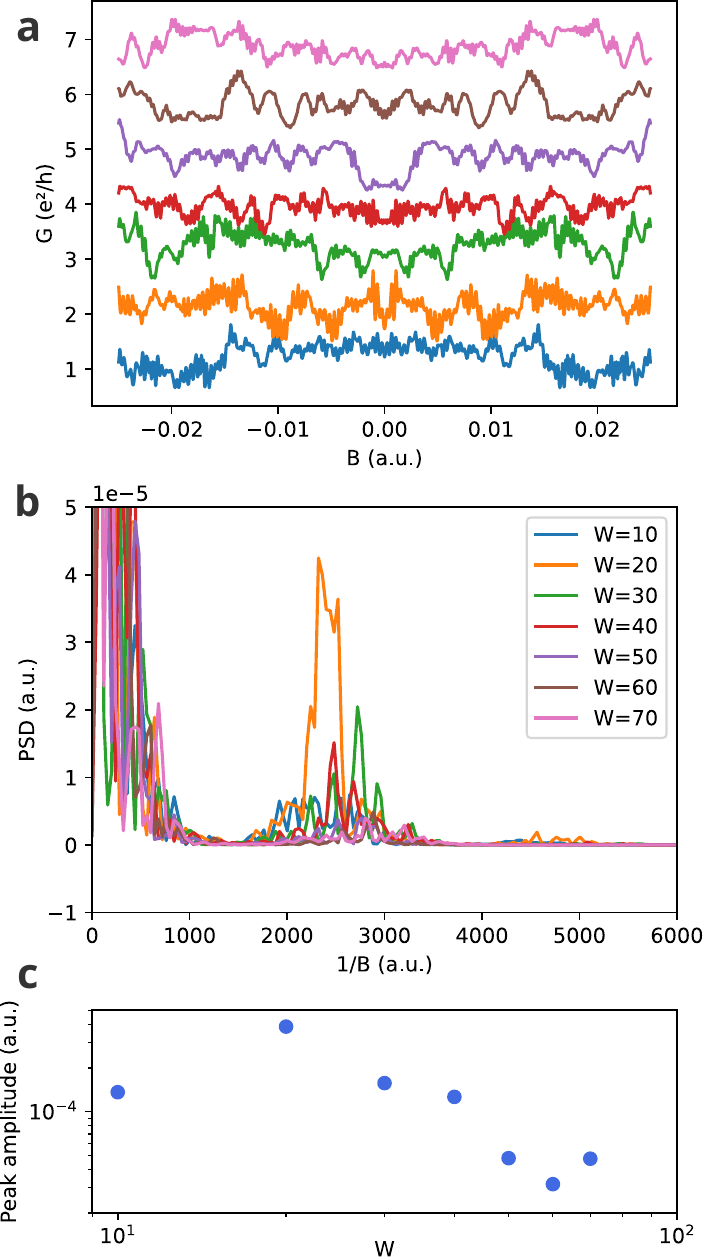}
		\caption{Simulation of the AB effect for different separation $W$ of ballistic edge states. The length $L$ of the ballistic channels is adjusted so as to keep the same periodicity of the AB effect. a) Magnetoconductance for different widths. The colors correspond to legend of b. (b) FFT of the magnetoconductance. (c) Amplitude of the FFT peak integrated between 1200 and 3700 for the different widths.}
		\label{fig:W_ABeffect}
    \end{figure}
    
    \begin{figure}[tb]
		\centering
		\includegraphics[width=\columnwidth]{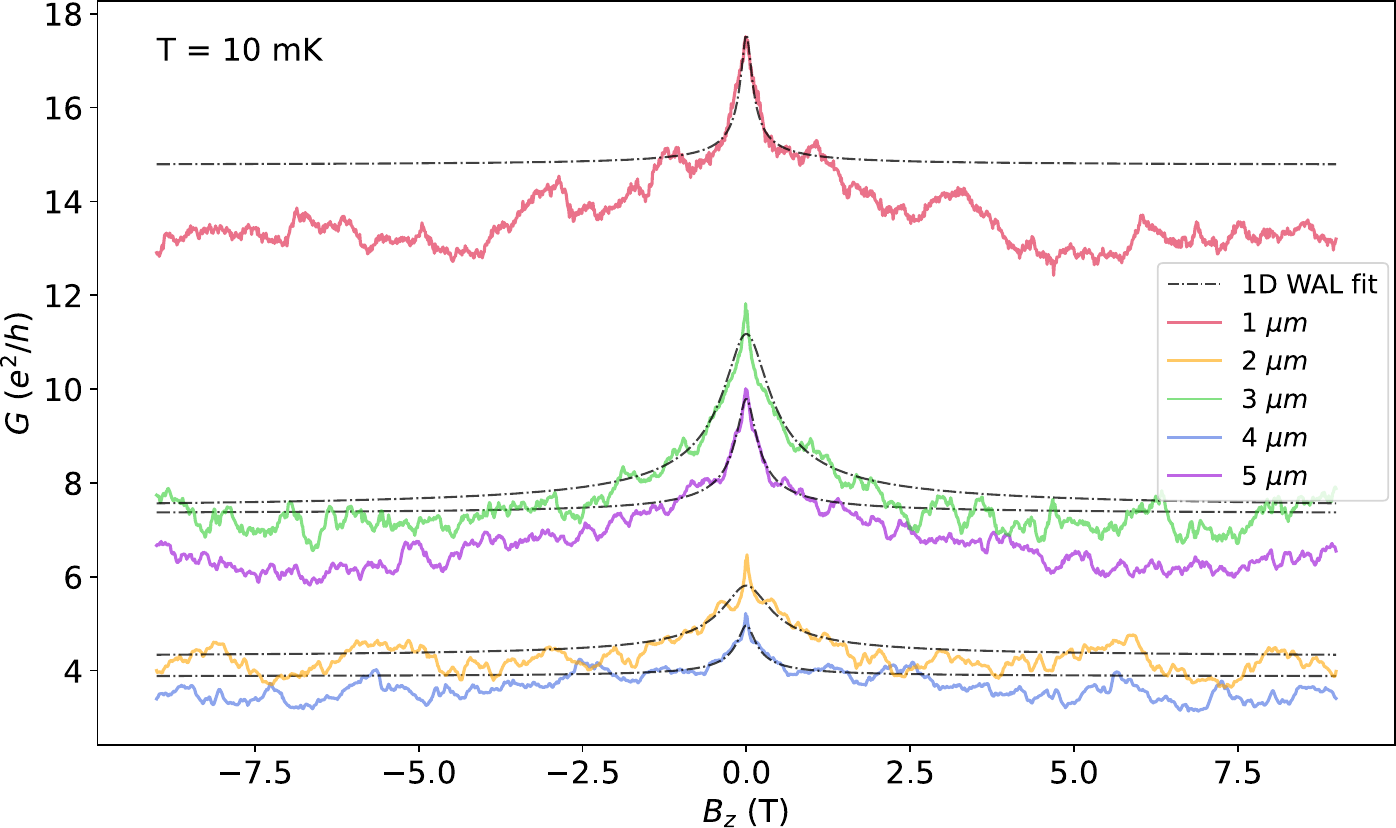}
		\caption{Fit of the magnetoconductance using formula \ref{eq:WAL1D} for quasi-1D geometry. The quality of the fit is not very good and lead to non consistent parameters.}
		\label{fig:WAL1D}
    \end{figure}
    
    \begin{figure}[tb]
		\centering
		\includegraphics[width=\columnwidth]{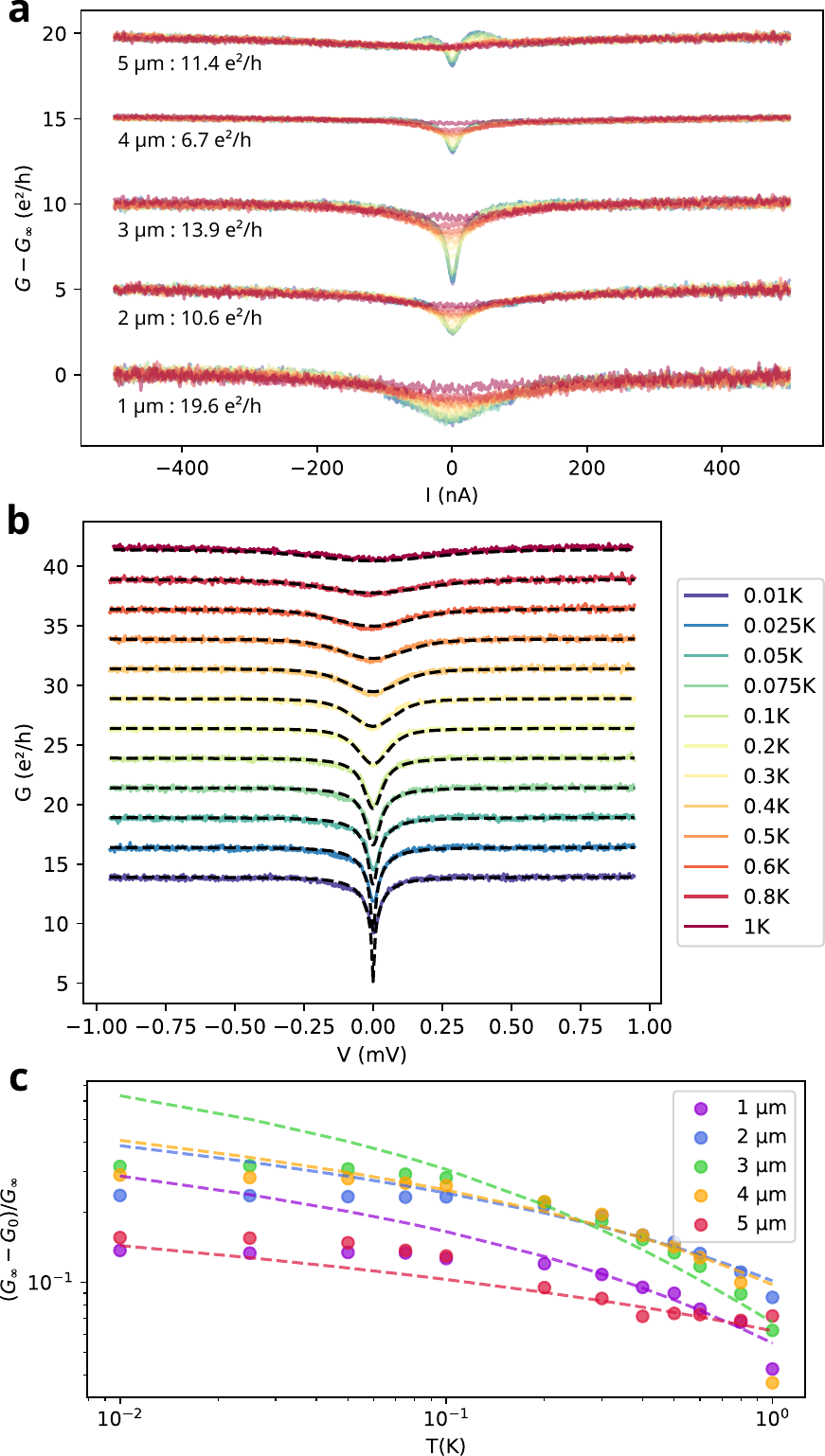}
		\caption{a) Differential conductance of the different segments of sample $\mathcal{L}$ as a function of DC current, for temperatures indicated by the legend of b. The length of the considered segment is indicated together with the high bias conductance $G_0$. The curves are shifted for clarity. b) Differential conductance $dI/dV$ of the 3$\mu$m segment of sample $\mathcal{L}$ and fit with the prediction of ref. \cite{Golubev2001} (eq. 25) for temperature indicated by the legend. c) Temperature dependence of the relative amplitude of the zero bias peak. All the peaks exhibit a temperature scale in the Kelvin range. The dashed lines indicates the temperature behaviour expected from the fit based on ref. \cite{Golubev2001} }
	    \label{fig:ZBA}
    \end{figure}
    
	\begin{figure}[tb]
		\centering
		\includegraphics[width=\columnwidth]{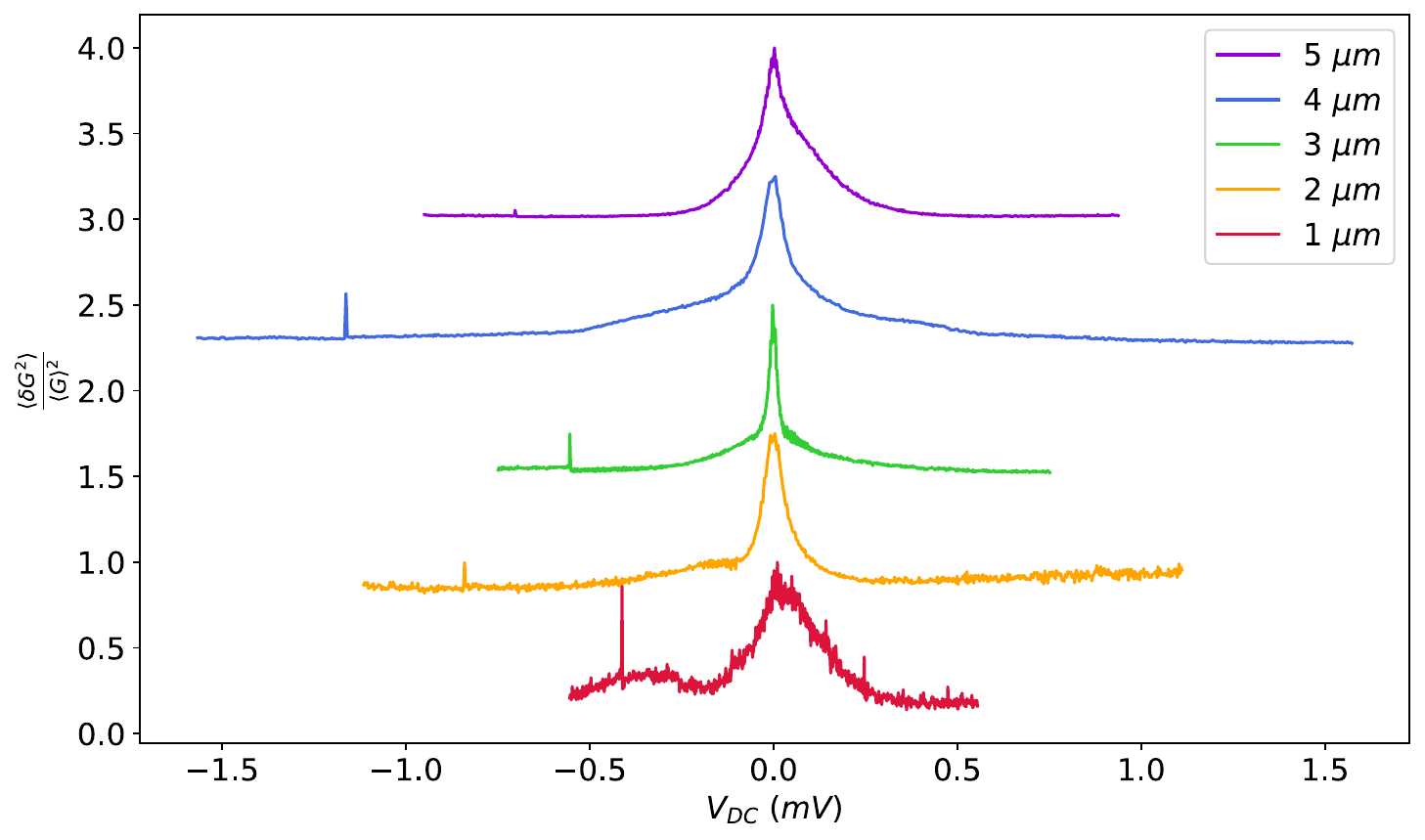}
		\caption{Ratio of the amplitude of the conductance fluctuation, deduced from the gate voltage dependence of the conductance, and conductance squared for the different segments of sample $\mathcal{L}$. The ratio is normalized by the value at zero bias. The curves are shifted by 0.75 for clarity.}
	    \label{fig:UCF_V_avg}
    \end{figure}
    
    \begin{figure}[tb]
		\centering
		\includegraphics[width=\columnwidth]{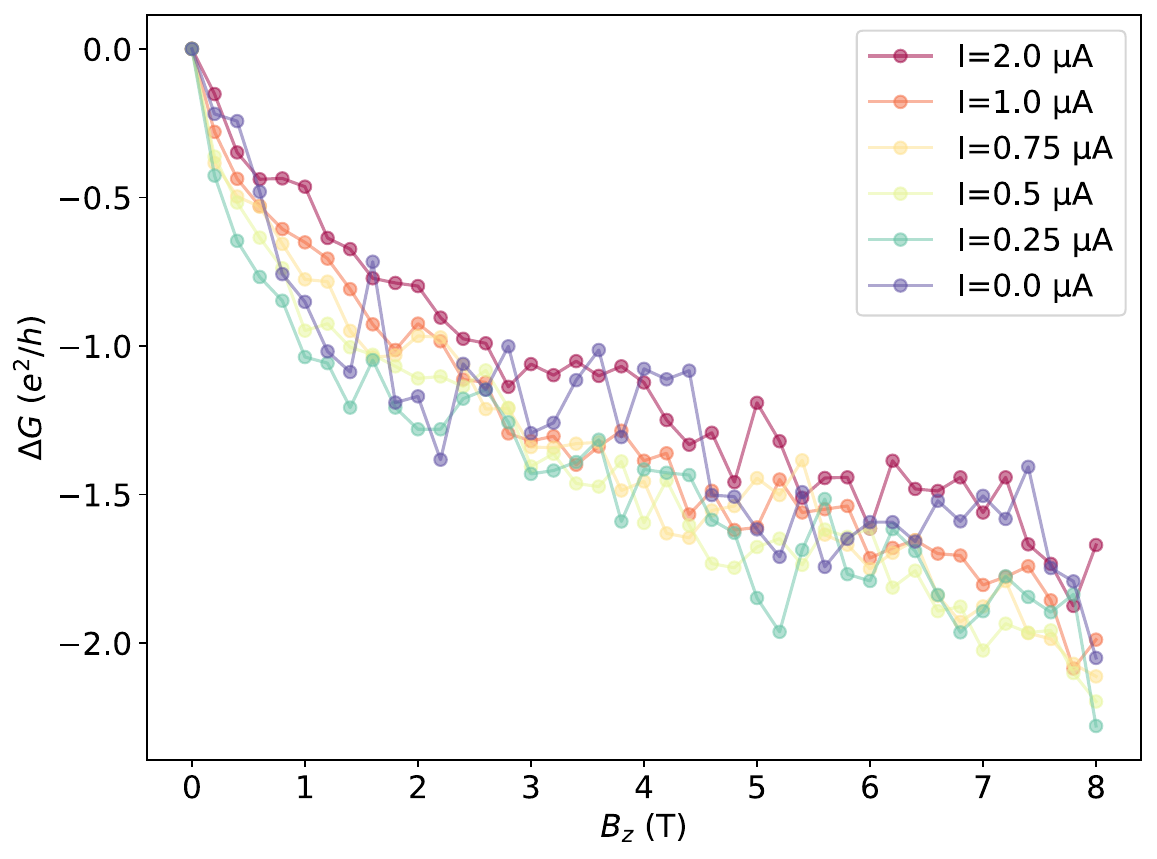}
		\caption{WAL correction $\Delta G = G(B)-G(B=0 \ T)$ for varying DC current biases at T = 10 mK.}
	    \label{fig:WAL-current}
    \end{figure}
    
    \begin{figure*}[tb]
		\centering
		\includegraphics[width=0.9\textwidth]{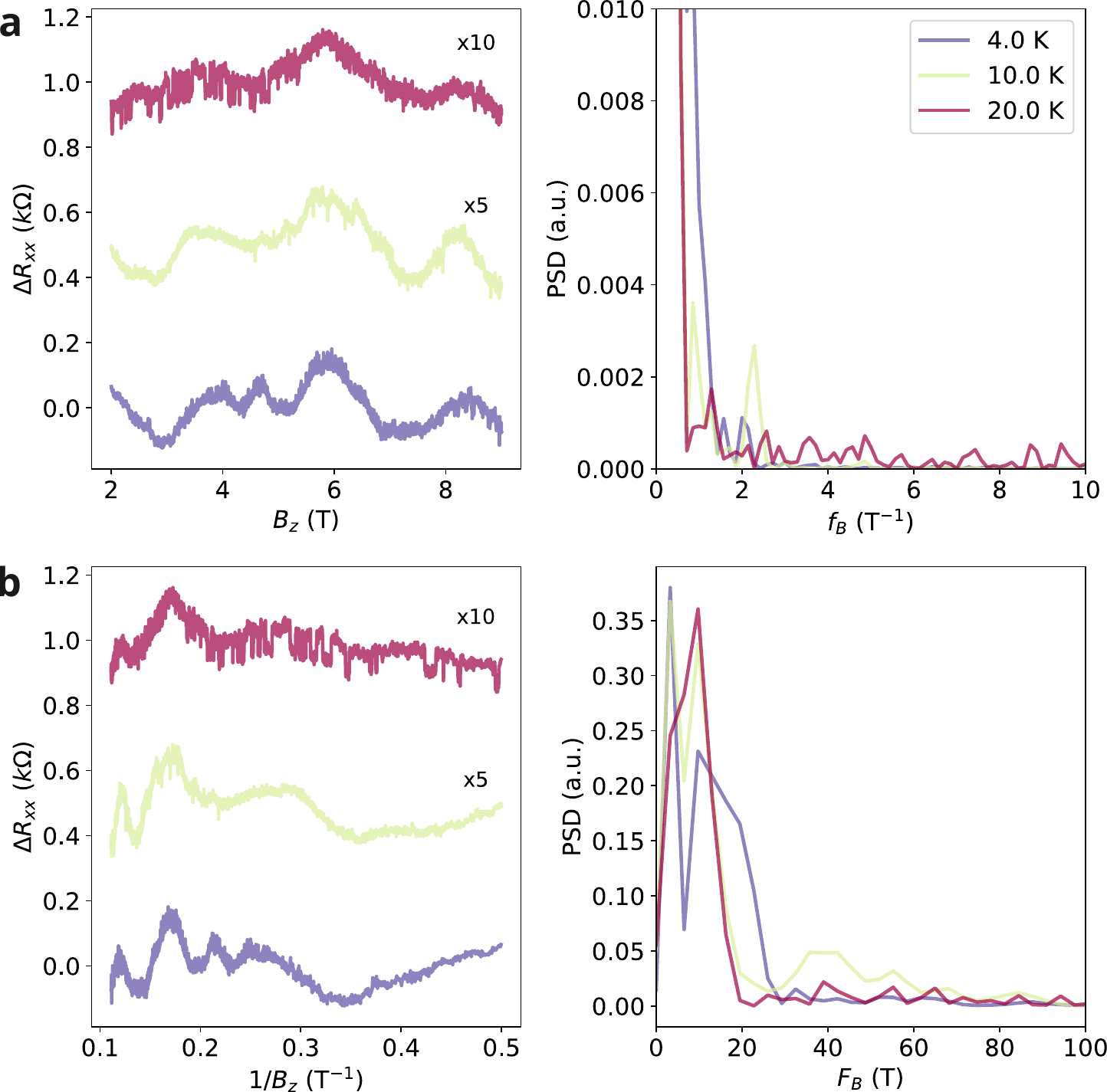}
		\caption{\color{black} a) Left: 4-wire longitudinal resistance of sample $\mathcal{H}_1$ as a function of out-of-plane magnetic field $B_z$ at temperatures 4 K, 10 K and 20 K (indicated by legend). A linear baseline has been removed, curves have been shifted by 500 $\Omega$ for clarity, and data at 10 K and 20 K are multiplied by a factor shown in the figure. Right : power spectral density (PSD) of the same data, showing no particular periodicity in the signal, as expected for the UCFs. b) Same data plotted as a function of 1/$B_z$. No specific periodicity appears in the direct signal or PSD.}
	    \label{fig:H1_SdH}
    \end{figure*}
    
%%%%%%%%%%%%%%%%%%%%%%%%%%% Cristal Quality %%%%%%%%%%%%%%%%%%%%%%%%%%%%%%%%
    \section{Crystal quality and contact region}
        As a reference, we assess the disorder in pristine exfoliated flakes before any contact is made. To this end, we exfoliate Bi$_4$Br$_4$ in a glove box using bare PDMS stamps. The Bi$_4$Br$_4$-covered stamps are brought into contact with a silicon wafer piece previously spin-coated with PMMA 950 A3. The chip is then pressed onto a Transmission Electron Microscope (TEM) copper grid (of size 1500 mesh), which sticks to the PMMA. It is then briefly taken outside the glove box to dissolve the PMMA in an acetone bath, thus separating the TEM grid from the wafer piece. This procedure results in the deposition of Bi$_4$Br$_4$ flakes onto the TEM grid, some of which bridge across the holes (Fig.\ref{fig:cristal_quality}a), making them suitable for observation. Images of several flakes revealed excellent crystalline quality, despite an exposure to ambient air of approximately one hour. An example is shown in (Fig.\ref{fig:cristal_quality}b and c), where the flake is aligned along the a direction. Strikingly, the flake maintains its crystalline order up to the outermost layers.
        
        The FIB cross-section of the contact regions have been observed and analyzed in a STEM (Microscope FEI ThermoFisher TEM/STEM Titan Themis 200). The areas under the contacts of samples with contacts on top show significant damage to the crystal structure of Bi$_4$Br$_4$, and the elemental maps reveal an increase in bismuth and palladium content, with a simultaneous decrease in bromine content directly below the gold (Fig. 1e of the main text). We attribute the degradation of Bi$_4$Br$_4$ into amorphous Bi to the IBE step before contact deposition, which causes interdiffusion of Bi in the thin layer of Pd deposited on top. The damaged areas can extend to several tens of nanometers away from the contacts.
    	
    	Even in the absence of a disorder-inducing step such as IBE, ohmic contacts made on Bi$_4$Br$_4$ are invasive. Indeed, we present STEM-EDX elemental maps on samples with contacts below the flakes (Fig. 1 f,g of the main text).  The STEM images reveal that this method does not result in a uniform interface between the flake and the Pd track. The flake is separated from the contact by a nearly continuous 3 nm-thick layer, pierced by asperities of the Pd electrodes, thereby permitting an electrical contact with the flake. Further inspection of this layer reveal that it is rich in silicon and carbon, hinting towards residues of uncrosslinked-PDMS which seem to cover nearly all of our Bi$_4$Br$_4$ flakes. This thin layer of residue likely comes from the exfoliation process, and may contribute to the relatively long stability in air that we have observed in our samples.
    	
    	It is interesting to note that the regions in which the residue layer has been pierced by Pd show Bi$_4$Br$_4$ crystal damage similar to that of the evaporated contacts. Amorphous regions show up to 25 \% of Pd and 65 \% Bi, indicating a diffusion of the palladium from the contacts into the crystal over several tens of nanometers. This interdiffusion process is somewhat anisotropic : the diffusion regions are generally less than 20 nm deep and extend laterally about 100nm.
        
%%%%%%%%%%%%%%%% Resistance network determination %%%%%%%%%%%%%%%%%%%%%%%%%%	
	\section{Conductance matrix for samples $\mathcal{H}_2$ and $\mathcal{H}_3$}
    	We present the conductance matrices measured on two extra samples, $\mathcal{H}_2$ and $\mathcal{H}_3$ measured at 4 K and 300 K respectively. For all samples the resistance values we find are all in the k$\Omega$ range (see main text and Fig \ref{fig:RmatH2H3}). These conductance matrices confirm the anisotropy discussed in the main text.  

%%%%%%%%%%%%%%%% Resistance and Length dependence %%%%%%%%%%%%%%%%%%%%%%%%%%		
	\section{Resistance and Length dependence of sample $\mathcal{L}$ and $\mathcal{L}_2$}
	
    	\subsection{Length dependence of resistance for sample $\mathcal{L}$}
        	We present the data for two Bi$_4$Br$_4$ samples ; $\mathcal{L}$ is contacted by Pd(7 nm)/Au(100 nm) electrodes onto a doped silicon substrate, which serves as a back gate, covered by 500 nm of oxide and  $\mathcal{L}_2$, which was stamped onto 20 nm thick palladium electrodes. Both were measured in a dilution fridge with a base temperature of around 10 mK. Sample $\mathcal{L}$ (Fig \ref{fig:A-B_amp} of main text) has five segments, ranging from 1 $\mu$m to 5 $\mu$m, and $\mathcal{L}_2$ has segments of length 200 nm to 1 µm, which resistances are reported in tables \ref{tab:res} and \ref{tab:res2} respectively. For both samples, the resistance of the segments do not increase linearly with their lengths, which indicates that they are dominated by a contact resistance.
        
            \begin{table}[!htb]\label{tab:res}
                \begin{center}
                    \begin{tabular}{| c | c | c | c | c | c |}
                    	\hline
                    	Length ($\mu$m) & $\quad$1$\quad$ & $\quad$2$\quad$ & $\quad$3$\quad$ & $\quad$4$\quad$ & $\quad$5$\quad$ \\
                    	\hline
                    	R (k$\Omega$) & 1.6 & 3.2 & 2.8 & 5.5 & 2.7\\
                    	\hline
                    \end{tabular}
                \end{center}
                \caption{Length and resistance of the different segments of sample $\mathcal{L}$.}
            \end{table}
            
            \begin{table}[!htb]\label{tab:res2}
                \begin{center}
                    \begin{tabular}{| c | c | c | c | c | c | c | c | c | c | c |}
                    	\hline
                    	Length (nm) & 200 & 300 & 400 & 500 & 600 & 700 & 800 & 900 & 1000\\
                    	\hline
                    	R (k$\Omega$) & 5.02 & 1.86 & 2.01 & 1.36 & 1.56 & 2.05 & 1.91 & 0.96 & 5.98\\
                    	\hline
                    \end{tabular}
                \end{center}
                \caption{Length and resistance of the different segments of sample $\mathcal{L}_2$.}
            \end{table}
        
        \subsection{Gate and field dependence of conductance fluctuations}
            As presented in the main text, sample $\mathcal{L}$ exhibits conductance fluctuations at low temperatures as a function of magnetic field. We found that these fluctuations also exist as a function of gate voltage, with a similar amplitude, as expected from the ergodic nature of UCF (Fig. \ref{fig:UCF_Vg}).

%%%%%% Temperature dependence of the Aharonov-Bohm effect %%%%%%%%%%%%%%%%%%%%%%
	\section{Aharonov-Bohm effect of samples $\mathcal{H}_1$ and $\mathcal{L}$}
	
	    {\color{black} The resistance of sample $\mathcal{H}_1$ exhibits AB oscillations. The period of these oscillations does not depend on temperature but the amplitude does (Fig. \ref{fig:AB_H1}) and leads to the exponential decay discussed in the main text.}
	    
	    The peaks identified on the FFT of the magnetoconductance of sample $\mathcal{L}$, denoted $\alpha$, $\beta$, $\gamma$ and $\delta$ (Fig. 9b and c of the main text), can be associated to different locations of the edge states. These locations corresponds to relatively simple stackings of a few Bi$_4$Br$_4$ unit cells (Inset of Fig. \ref{fig:A-B_amp}), yielding reasonable agreement between the measured and expected values for the magnetic frequency rescaled to a 1$\mu$m segment $f_B^{1\mu\mathrm{m}}$ (Table \ref{tab:AB_L1}). For the $\gamma$ peak, the large difference in height (7.9 nm) between the edge states in the proposed stack allows it to be tested with a transverse magnetic field in the range achievable with our vector magnet. The corresponding experimental data is shown in Fig. \ref{fig:A-B_L1_By} and indeed exhibits a consistent peak in the Fourier transform at the expected magnetic frequency ($f_B/L$=1.92$\mathrm{T}^{-1}$). 
	    
	    As for the temperature dependence of sample $\mathcal{H}_1$, we find a exponential decay with temperature for a majority of identified AB peaks in the FFT of all segments of sample $\mathcal{L}$. This dependence is calculated by measuring the area under an identified peak ($\alpha$, $\beta$, $\gamma$ and $\delta$, and their harmonics) as a function of temperature. Fitting of this area to $a e^{-bT}$ is done for the first harmonic, the n$^\mathrm{th}$ harmonics are allowed only a 5\% adjustment around $a / n$ and $b \times n$, which gives a reasonable agreement for a smaller set of peaks. This exponential decay is consistent with a $1/T$ dependence of the phase coherence length and a value of this length at 1K between 1.5 and 6 $\mu$m, of the same order of magnitude than the scale obtained on sample $\mathcal{H}_1$. 
    	
    	\begin{table}[!htb]
				\begin{center}
				    \begin{tabular}{|c|c|c|c|c|}
                        \hline
                        $\quad$Peak label$\quad$ & $\quad\alpha\quad$ & $\quad\beta\quad$ & $\quad\gamma\quad$ & $\quad\delta\quad$ \\
                        \hline
                        $f_B^{1\mu\mathrm{m}}$ ($\mathrm{T}^{-1}$) & 0.61 & 0.77 & 1 & 1.4 \\
                        $W^z_\mathrm{meas}$ (nm) & 2.5 & 3.2 & 4.1 & 5.8 \\
                        $W^z_\mathrm{th}$ (nm) & 2.61 & 3.23 & 4.07 & 5.84 \\
                        $W^y_\mathrm{th}$ (nm) & 0 & 2.6 & 7.9  & 2.6 \\
                        \hline
				    \end{tabular}
				\end{center}
				\caption{Modelization of the measured magnetic frequencies $\alpha - \gamma$ in the Fourier transform of sample $\mathcal{L}$ with simple stackings of a few Bi$_4$Br$_4$ unit cells (Fig. \ref{fig:A-B_amp}). For each peaks the measured magnetic frequency $f_B^{1  \mu\mathrm{m}}$, the corresponding width $W^z_\mathrm{meas}$ and the associated horizontal ($W^z_\mathrm{th}$)  and vertical ($W^y_\mathrm{th}$) distance are given.}
			    \label{tab:AB_L1}
		\end{table}

%%%%%%%%%% Sensitivity of H1 to disorder %%%%%%%%%%%%%%%%%%%%%%	

        Sample $\mathcal{H}_1$ has proven to be quite sensitive to small current spikes, that seem to have modified its disorder each time it has happened. As a consequence, we have selected only a few curves to show in the main text.
        
        First, we take a look at the reproducibility of AB oscillation with variations in disorder. We show five curves, corresponding to three disorder realizations and different current biases. We can first notice a strong reduction of the AB amplitude when increasing the AC current from 1 nA to 10 nA. This seems to be linked to the nonlinearity of the conductance of Bi$_4$Br$_4$ sections, as discussed in the section "Zero bias anomaly". Hence, we stuck to 1 nA current bias to get the maximum AB amplitude. Nevertheless, all of our data show a pronounced peak in Fourier space $f_B=4.16$ T$^{-1}$, indicating its intrinsic nature. Because the study of AB oscillations with varying temperature was only done for configuration $\mathbf{C}$, it is the data shown in the main text.
        
        Then, we examine the consistence of the observed negative 4-wire resistance. We see that all of the longitudinal magnetoresistance traces taken with the two top electrodes exhibit negative values in some field ranges. However, we could not benefit from the increased AB amplitude at lower biases, due to spurious time-dependent telegraphic noise hindering the repeatability of our data at 1 nA.
        Interestingly, the four-wire resistance measured with the two bottom electrodes (not shown) does not exhibit excursions to negative values. This may be explained by a difference in invasiveness between the two sets of electrodes ; the two-wire resistance through the top electrodes is $\sim 130 \ k\Omega$, while it is $\sim 35 \ k\Omega$ for the bottom ones.
        
%%%%%%%%%% Numerical simulation of the UCF of a 1D channel %%%%%%%%%%%%%%%%%%%%%%	
	\section{Numerical simulation of the UCF and AB effect of ballistic channels}
	
	Transport simulations shown on figure 8 of the main text were made using Kwant Python package \cite{Groth2014}. We have simulated ballistic wires (200x3 sites) between two diffusive regions (50x100 sites) on a square lattice of parameter $a$ (see insets of Fig. 8), with one orbital and no spin degree of freedom, using the following tight-binding Hamiltonian :
     \begin{equation}
     	\hat{\mathbf{H}} = \sum_{i}\left(4t + D_{\mathrm{rand}} \right)a_i^\dagger a_i + \sum_{\langle i;j\rangle}  t_{ij}(\phi) a_i^\dagger a_j
     \end{equation}
 	The onsite Hamiltonian has an added disorder term $D_{\mathrm{rand}}$ which equals 0 in the ballistic wires, and follow a uniform random distribution over $[-D,D]$ elsewhere. The out-of-plane magnetic field is incorporated in the hopping term using the Peierls substitution, with :
 	 \begin{equation}
        t_{ij}(\phi) = - t \exp{\left(-i \phi(x_i - x_j)(y_i + y_j)/2\right)}
     \end{equation}
 	The simulation shown in the main text were made using the following parameters : $a=1$, $t = 1$, $D=1$, $\phi \in [-0.05;0.05]$, $E = 1.8$. We simulate two geometries. The first one is a single ballistic wire between two diffusive regions, which exhibits conductance fluctuations (Fig. 8a of the main text). The second configuration corresponds to two ballistic wires. In this case the conductance exhibits AB oscillation with a field scale determined by a flux quantum in the area enclosed by the two wires (Fig. 8b of the main text).
 	
     The question arises as to whether trajectories in the region of diffusive contacts could blur the Aharonov-Bohm effect. This is for example what is seen in AB ring with low aspect ratio (large width compared to perimeter). However we find on the contrary that the AB effect, especially for sample H1 (Fig. 7 of the main text) has a relatively well defined magnetic field period that seems in contradiction with the previous point. We think that this good visibility of the AB oscillations is due to the fact that the interfering 1D edge states are separated by a distance of the order of the mean free path of the diffusive region. In this limit the two paths cannot be considered independently in a quasi-classical picture and lead to relatively well define AB oscillations. When the separation of the edge states is much higher than the mean free path, the contribution from different trajectories in the region of diffusive contacts leads to a strong decrease of the visibility of the oscillation. We have conducted numerical simulations using Kwant package to support this point. For this we have considered two 1D ballistic states separated by a distance W and contacted to two diffusive coherent contacts (100x100 sites, disorder=2.0). We have calculated the conductance as a function of magnetic field for separation varying between 10 and 70 lattice steps (Fig. \ref{fig:W_ABeffect}). With the chosen disorder strength $w$ and energy $E$ the mean free path $l_e=48a \sqrt{E/t} (t/w)^2$ is estimated to 16 lattice sites. We find that the amplitude of the AB effect is relatively high for separation of the order of $l_e$. However when the separation is higher than several $l_e$ the AB effect is much less visible. This expected blurring of the AB oscillations should occur in rings connected with constrictions, but such experiments likewise show well-defined magnetic field periodicities \cite{Wiel2003,Yamauchi2009}.
     
%%%%%%%%%% WAL 1D %%%%%%%%%%%%%%%%%%%%%%%%%%%%%%%%%%%%%%%%%%%%%%%%%%%%%%%%%%%%%%%
	\section{Fitting of magneto conductance with 1D weak anti-localization}
	
    	We have tried to fit the magnetoconductance of sample $\mathcal{L}$ with the formula describing weak anti-localization for quasi-1D geometry using formula :
    	\begin{equation}
    		\delta G(B) = \frac{\alpha e^2}{2 h L}\left( \frac{1}{l_\varphi^2} 
    		+ \frac{e^2 B^2 W^2}{3 \hbar^2} \right)^{-1/2}
    		\label{eq:WAL1D}
    	\end{equation}
    	with $l_\varphi$ the phase coherence length, $L$ the length of the sample, $W$ its width, $B$ the magnetic field and  $\alpha$ a prefactor corresponding to the number of independent channels \cite{Montambaux2007}. The fitted data are shown on figure \ref{fig:WAL1D}.
	    
	    Assuming that the lengh $L$ is the one of the considered segment, one obtains fit parameters listed in Table \ref{tab:1DWAL}. The parameters are not consistent and vary a lot from segment to segment. 
	    \begin{table}[!htb]
            \begin{center}
            \caption{Fit parameters using 1D WAL for sample $\mathcal{L}$.}
            \label{tab:1DWAL}
                \begin{tabular}{| c | c | c | c | c |}
                	\hline
                	Length ($\mu$m) & $\alpha$ & $l_\varphi$ (nm) & W (nm) & $G_0$ ($e^2/h$) \\
                	\hline
                	5 & 20.9 & 586 & 13.6 & 7.33 \\
                	4 & 37.7 & 116 & 76 & 3.87 \\
                	3 & 12.8 & 878 & 4 & 7.43 \\
                	2 & 20 & 153 & 22 & 4.28 \\
                	1 & 1.1 & 2429 & 6 & 14.76\\
                	\hline
                \end{tabular}
            \end{center}
        \end{table}

%%%%%%%%%% ZBA %%%%%%%%%%%%%%%%%%%%%%%%%%%%%%%%%%%%%%%%%%%%%%%%%%%%%%%%%%%%%%%	   
	\section{Zero bias anomaly}
	    The features (WAL, UCF and AB effect) described in the main text of the article concern the linear regime of quantum transport. When applying a DC current bias through the sample, added to the small ac current (typically in the nA range) used to probe the differential resistance, this latter quantity exhibits a peak around zero voltage (Fig. \ref{fig:ZBA}a). This is a universal characteristic for all our samples, even if the amplitude and width of the peak vary from sample to sample. 
	    The zero bias anomaly (ZBA) has been reported in very different systems spanning from tunnel junctions \cite{Devoret1990,Ingold1992} or quantum point contacts \cite{Parmentier2011} in a resistive environment to transport in 1D systems as carbon nanotubes \cite{Bockrath1999,Gao2004}, edge states of quantum spin hall \cite{Li2015,Fei2017,Stuhler2020} or second order topological insulator \cite{Wang2023} and are either attributed to Coulomb interaction affecting quantum transport or Tomonaga Luttinger Liquid (TLL) physics \cite{Kane1992,Hsu2021}. Both effect can lead to power law dependence of ZBA \cite{Safi2004,DCB_exponent} and scaling behaviour but from different physical origin. Whereas the energy scale for DCB is determined by the charging energy $E_C$ of the conductor and the exponent $\alpha$ by the impedance of the electromagnetic environment and of the sample \cite{Ingold1992,Golubev2001,Yeyati2001,Kindermann2003,Golubev2005,DCB_exponent} , the TLL model cutoff is given by the Thouless energy $h v_F/L$ of the 1D conductor of length $L$ and Fermi velocity $v_F$ and the exponent $\alpha = (K+K^{-1}-2)/2 $ by the interaction parameter $K$.   
	    
	    The TLL scenario is appealing because of the expected 1D nature of the edge states in Bi$_4$Br$_4$. However, for sample $\mathcal{L}$, with section length $L$ varying from 1 to 5$\mu$m and $v_F$ around 5.10$^5$m.s$^{-1}$  \cite{Zhou2015,Yoon2020,Shumiya2022} one expects a power law behaviour for T higher than a cutoff temperature $\hbar v_F/(k_B L)$ \cite{Kane1992} between 0.8 and 3.8K. As a consequence we can exclude the TLL scenario since in all our samples, we have seen the ZBA disappear around 2K. Moreover this scenario should lead to a systematic variation with the length of the sample, which is not seen in the experiment. 
	    
    	Another possibility is that quantum transport is affected by Coulomb interaction. In this picture we assume that the samples can be described by several coherent conducting channels in parallel, and that the transport of each channel is affected by the back-action of the other channels. We note that if each channel is perfectly ballistic no effect of the electromagnetic environment should exist. However due to the existence of small diffusive coherent region between the contact and the Bi$_4$Br$_4$ crystal, we rather assume, based on results discussed in the main text of the article, that the samples are well described by imperfectly transmitted coherent channels, the transmission of which can be modulated by quantum interference effects. In this situation, Coulomb interaction leads to an energy dependence of the transmission coefficient and thus to ZBA \cite{Ingold1992,Golubev2001,Yeyati2001,Kindermann2003,Safi2004,Golubev2005}. We apply the results for Coulomb blockade in a coherent conductor due to its own resistance \cite{Golubev2001} and use in particular the relation giving the current voltage characteristic of the conductor:
    	\begin{equation}
            I=G_0 V-\frac{e \beta k_B T}{\hbar} \mathrm{Im}\left[ w \Psi \left(1+\frac{w}{2}\right) -i v \Psi\left(1+i \frac{v}{2}\right) \right]
            \label{eq:ZBA}
        \end{equation}
        with $w=u+i v$, $u=h G_0 E_C / (e^2 \pi^2 k_B T)$, $v=eV/(\pi k_B T)$ and $\Psi$ the digamma function. In this equation the adjustable parameters are $G_0$, the conductance of the sample without interaction effect, $\beta$, an effective Fano factor which is associated with the transmission of the conductor considered and gives the amplitude of the effect at very low temperature, and $E_C$ the charging energy.

    	Figure \ref{fig:ZBA}b shows the fitted curves for the 3 $\mu$m long segment of sample $\mathcal{L}$ at different temperatures. We give in table \ref{Table:ZBA_fit} the fitted parameters for each segment.
	
    	 \begin{table}[!htb]
            \begin{center}
                \label{Table:ZBA_fit}
                \begin{tabular}{|c | c | c | c|} 
                     \hline
                     Length ($\mu$m) & $E_C$ (meV) & $\beta$ & $G_0$ ($e^2$/h) \\ 
                     \hline\hline
                     1 & 0.041 & 0.52 & 19.59 \\ 
                     \hline
                     2 & 0.155 & 0.33 & 10.62 \\
                     \hline
                     3 & 0.020 & 0.99 & 13.90 \\
                     \hline
                     4 & 0.196 & 0.23 & 6.72 \\
                     \hline
                     5 & 0.99 & 0.10 & 11.41 \\
                     \hline
                \end{tabular}
                \caption{Length and ZBA fit parameters (see text) for the different segments of sample $\mathcal{L}$.}
            \end{center}
        \end{table}
    	
    	This agreement is only consistent for temperatures higher than 80 mK, that we attribute to a saturation of electronic temperature below this limit. We note however that formula \ref{eq:ZBA} cannot explain the peaks in dI/dV seen in the longer segment (Fig. \ref{fig:ZBA}a). We postulate that this may be due to the intrinsic energy dependence of the transmission coefficients. Furthermore, the relatively high dispersion of the parameters we obtain for the different segments calls for a better understanding of this non-linear effect.

	In parallel with the ZBA, coherent effects disappear at finite current. This affects the conductance fluctuation on a voltage scale of the order of 50 $\mu$eV (deduced from half width at half maximum), with no clear correlation with the length of the section considered (Fig. \ref{fig:UCF_V_avg}). However the WAL correction is not affected by a finite bias (Fig. \ref{fig:WAL-current}). This rules out thermal effects and self-heating that would otherwise affects both UCF and WAL through the temperature dependence of the phase coherence length. The energy scale affecting the amplitude of the UCF is expected to be related to the Thouless energy in diffusive coherent sample \cite{Altshuler1985,Khmelnitskii1986,Webb1988} but can involve also the mean level spacing \cite{Tang1991}. In our case the Thouless energy $E_{th}^{1D}$of the 1D helical states $\hbar v_F/L$ gives an energy scale between 330 $\mu$eV for the shorter section and 66 $\mu$eV for the longer one, taking $v_F=5 \, 10^5 \ m.s^{-1}$. This is not consistent with the voltage scale measured especially because of the absence of correlation with the length of the sample. The Thouless energy $E_{th}^{D}$ of the diffusive coherent contact can be estimated $E_{th}^{diff}= \hbar D/L_D^2$ using $D=v_F^D l_e^D/3$ where we take reasonable values for the unknown Fermi velocity of the Bi/Pd alloy $v_F^D=10^6 \  m.s^{-1}$ and its mean free path $l_e^D=5$ nm, leading to $E_{th}^{D}=110 \ \mu$eV for a length $L_D$ of the diffusive region of 100 nm. So it seems that the measured voltage scale of the UCF decrease is more consistent with the one imposed by the coherent diffusive region.
	
	 \section{Absence of Hall effect or Shubnikov de Haas oscillations}
	 
	 As detailed in section 3 of the main text, sample $\mathcal{H}_1$ does not show any Hall effect. This is consistent with the very strong anisotropy of electronic transport in this sample, which together suggest the absence of bulk and (001) surface conduction. Figure \ref{fig:H1_SdH} shows the magnetoresistance of the same sample for temperatures between 4 K and 20 K as a function of out-of-plane magnetic field $B_z$ and 1/$B_z$ for a field greater than 2 T and the corresponding power spectral density. This analysis shows that the conductance fluctuations has no specific periodicity as expected for UCFs, thus are not consistent with Shubnikov-de Haas oscillations nor Aharonov-Bohm oscillations.

%%%%%%%%%% Biblio %%%%%%%%%%%%%%%%%%%%%%%%%%%%%%%%%%%%%%%%%%%%%%%%%%%%%%%%%%%%%%%